\documentclass[12pt,a4paper]{article}
\usepackage{amsmath,amssymb,amsthm,amsfonts}
\usepackage{breqn}
\usepackage[margin=1.0in,rmargin=1.in,tmargin=2cm,bmargin=3cm]{geometry}
\usepackage{cite}
\usepackage{graphicx}
\usepackage[pagebackref=false]{hyperref}
\usepackage[affil-it]{authblk}
\numberwithin{equation}{section}
\def\a{\alpha}
\def\b{\beta}
\def\g{\gamma}
\def\d{\delta}

\def\k{\kappa}
\def\l{\lambda}
\def\m{\mu}
\def\n{\nu}

\def\s{\sigma}
\def\t{\tau}

\def\c{\chi}

\def\S{\Sigma}
\def\U{\Upsilon}

\def\L{\Lambda}

\newcommand{\nn}{\nonumber}

\def\be{\begin{equation}}
\def\ee{\end{equation}}
\def\bea{\begin{eqnarray}}
\def\eea{\end{eqnarray}}
\def\bg{\begin{align}}
\def\eg{\end{align}}

\def\pa{\partial}

\makeatletter
\renewcommand\section{\@startsection {section}{1}{\z@}%
                                   {-3.5ex \@plus -1ex \@minus -.2ex}
                                   {2.3ex \@plus.2ex}%
                                   {\normalfont\large\bfseries}}
\renewcommand\subsection{\@startsection{subsection}{2}{\z@}%
                                     {-3.25ex\@plus -1ex \@minus -.2ex}%
                                     {1.5ex \@plus .2ex}%
                                     {\normalfont\bfseries}}
\makeatother

\begin{document}
\begin{titlepage}
{\title{{Ricci cubic gravity in d dimensions, gravitons and SAdS/Lifshitz black holes }}}
\vspace{.5cm}
\author{Ahmad Ghodsi \thanks{a-ghodsi@ferdowsi.um.ac.ir}}
\author{Farzaneh Najafi \thanks{fa.najafi@mail.um.ac.ir}}
\vspace{.5cm}
\affil{ Department of Physics, Ferdowsi University of Mashhad,    
\hspace{5.5cm} P.O.Box 1436, Mashhad, Iran}
\renewcommand\Authands{ and }
\maketitle
\vspace{-12cm}
\begin{flushright}
\end{flushright}
\vspace{10cm}
\begin{abstract}
A special class of higher curvature theories of gravity, Ricci Cubic Gravity (RCG), in general $d$ dimensional space-time has been investigated in this paper. We have used two different approaches, the linearized equations of motion and auxiliary field formalism to study the massive and massless graviton propagating modes of the AdS background. Using the auxiliary field formalism, we have found the renormalized boundary stress tensor to compute the mass of  Schwarzschild AdS and Lifshitz black holes in RCG theory. 
\end{abstract}
\end{titlepage}
\section{Introduction}
Einstein-Hilbert action, as an effective gravitational theory, acquires different higher curvature corrections. The origin of these corrections may come from quantum gravity or string theory \cite{Stelle:1976gc, Zwiebach:1985uq, Metsaev:1986yb}. Specifically these gravitational theories with higher curvature corrections in presence of cosmological parameter become more important in the context of AdS/CFT correspondence (see for example \cite{Buchel:2009sk, Myers:2010tj}). 

There are many questions arising in these theories when one studies different black hole solutions. For example the existence of Schwarzschild $AdS$ ($SAdS$) or Lifshitz black hole is expected in these theories and consequently computation of mass or thermodynamical properties such as entropy will be a challenging problem. 

The linear excitation of the gravitational field or graviton mode is another important object in these theories. It is a well-known property for these theories to have  massive excitation modes in addition to the massless gravitons. The stability of vacuum solution requires tachyon-free conditions, which restrict the theory to  specific regions of the parameter space. 

Another common property in gravitational theories with higher curvature terms, is the existence of the scalar and tensor ghost modes. In pure theories of gravity although a scalar ghost mode can be eliminated by proper assumptions such as trace-less condition of the linearized equations of motion, but the tensor ghost modes may survive and destroy the unitarity of the dual CFTs. 
At first sight, the absence of tensor ghost modes can be achieved by going to the critical points, but at these points the massive modes degenerate into massless graviton mode and replace by ghost-like logarithmic modes. This theory may include a unitary subspace through the truncation of the logarithmic modes by imposing proper boundary conditions at the linear level. The unitarity problem of these theories has been discussed in various works \cite{Bergshoeff:2012sc,Bergshoeff:2012ev,Nutma:2012ss,Kleinschmidt:2012rs,Lu:2013hx, Apolo:2012vv}.

Many different properties of higher curvature theories of gravity have been investigated in different space-time dimensions. For example in $d=3$, gravitational theories known as massive gravities, have been studied extensively \cite{Deser:1982ac,Li:2008dq,Bergshoeff:2009hq, Bergshoeff:2009aq, Bergshoeff:2012ev, Afshar:2014ffa, Sinha:2010ai,  Paulos:2010ke, Ghodsi:2010gk, Ghodsi:2011ua, Ghodsi:2012fg}. Other higher curvature theories of gravity are also studied in five and six dimensions for example see \cite{Myers:2010ru} and \cite{Lu:2011ks}. In general $d$ dimensions one can follow several recent works, for example \cite{Gullu:2009vy, Oliva:2010zd, Sisman:2011gz, Bueno:2016xff, Bueno:2016ypa}.

In this paper we are interested to study a special class of higher curvature theories of gravity, Ricci Cubic Gravity (RCG), in general $d$ dimensional space-time in the context that we mentioned above. We will employ two different approaches, the linearized equations of motion and auxiliary field formalism. 

In first approach we study the linear excitations around a $d$ dimensional Anti-de Sitter ($AdS_{d}$) space-time  and we find the stability conditions of this black hole.
 We will show that there are restrictions on the free parameters of the RCG  when we are eliminating the scalar ghost modes. We also show that this model includes two massive graviton propagators and a massless one. We analyze various critical points of this theory where the massive modes are degenerated with the massless mode. We also compute the energy of excitation modes and the Abbott-Deser \cite{Abbott:1981ff} energy of different black hole solutions.

In second approach by a reformulation of RCG with the help of auxiliary fields we will find a Lagrangian of second order in derivatives of the fields.  The linearization  around the $AdS_{d}$ background up to the second order of gravitational coupling,  generates the Fierz-Pauli massive action. We can read again the mass of excitation modes by this approach.
 
The mass of  $SAdS$ and Lifshitz black holes can be computed in different ways, either by calculating the free energy and using the first law of thermodynamics or by computing the renormalized boundary energy-momentum tensor.

This paper is organized as follow: 
In section 2, we begin with a six derivatives action constructed out of Ricci curvature tensor and its covariant derivatives. We study the graviton modes by linearizing the equations of motion around the $AdS_d$ vacuum. In order to construct a theory free of the scalar ghost modes, we should impose two constraints on the couplings of this theory. We show that RCG contains two massive graviton modes in addition to a massless one. We also discuss about the stability of this vacuum solution. In last part of this section we calculate the conserved quantities of theory by the Abbott-Deser method \cite{Deser:2002rt}.

In Section 3, we reformulate the RCG action with the help of two auxiliary fields and we linearize it around the $AdS_{d}$ background up to the second order of gravitational coupling. Then we rewrite this action as a linear combination of three Fierz-Pauli massive Lagrangians for spin-two fields \cite{Bergshoeff:2012ev}.  We also find the energy of the linear excitations to reconfirm the stability arguments in section two.

In section 4, we will use the reformulated RCG action to compute the boundary energy-momentum tensor by using the technique which has been  introduced in \cite{Hohm:2010jc}. For this purpose, we will require a well-posed variational principle which provides by some generalized Gibbons-Hawking terms.

We study $SAdS$ black hole solution of RCG in section 5. We find the thermodynamical properties, such as temperature, free energy and entropy. We also compute the finite value of the mass of $SAdS$ from the renormalized boundary energy-momentum tensor by adding a proper counter-term to the boundary terms. We show that this mass is compatible with the first law of thermodynamics for black holes.

As a more complicated case, the Lifshitz black hole has been investigated in section 6. We have tried to find a finite mass from the boundary stress tensor, consistent with the first law of thermodynamics. We observe that similar to the three dimensional case in \cite{Hohm:2010jc} there is an ambiguity for writing the counter-terms.

Section 7 contains the results of previous sections but in special dimension $d=3$ to obtain the central charges of dual CFTs. In  last section we summarize and discuss about our results. 
Almost all parts of the calculations in this paper have been done by the Mathematica package xAct \cite{Nutma:2013zea}.

\section{Ricci cubic gravity in d dimensions}

Let us start with the most general Ricci Cubic Gravity (RCG) in $d$ dimensions by adding  all possible independent contractions of the Ricci tensor and its covariant derivatives to the Einstein-Hilbert action in the presence of a cosmological parameter $\Lambda_{0}$. We restrict ourselves to terms with at most six derivatives i.e.
\begin{align}\label{action}
S=\frac{1}{\kappa^{2}}\int d^{d}x \sqrt{-g} &\Big (\sigma R -2 \Lambda_{0}  +  a_{1} R^{\mu \nu} R_{\mu \nu} +a_{2} R^2
+b_{1} \nabla_{\mu}{R} \nabla^{\mu}{R} +b_{2} \nabla_{\mu}{R_{\alpha \beta}} \nabla^{\mu}{R^{\alpha \beta}}\nn\\
&+ c_{1} R^{\alpha \beta} {R_{\alpha}}^{\gamma} R_{\beta \gamma} + c_{2} R R^{\alpha \beta} R_{\alpha \beta}+ c_{3} R^{3}
\Big) \,,
\end{align}
where $\sigma$ is a dimensionless parameter and $\k$ is the gravitational coupling constant. These parameters together with the other couplings $a_{1}, a_{2}, b_{1}, b_{2}, c_{1}, c_{2}$ and $c_{3}$  make the parameter space of this theory.
The six-derivative equations of motion for action (\ref{action}) are given by 
\bea\label{EOM}
\sigma (R_{\mu \nu}-\frac12 R g_{\m\n}) + \Lambda_0 g_{\mu \nu} +  \sum_{i=1}^7 H^{(i)}_{\mu \nu}=0\,,
\eea
where
\begin{dgroup}
	\begin{dmath}\label{EOM1}
		H_{\mu \nu}^{(1)}= a_{1} \Big( (g_{\mu \nu} \Box \!-\! \nabla_{\mu} \nabla_{\nu}) R + \Box (R_{\mu \nu} - \frac{1}{2} R g_{\mu \nu}) +2 ( R_{\mu \alpha \nu \beta} -\frac{1}{4} g_{\mu \nu} R_{\alpha \beta}) R^{\alpha \beta} \Big)\,,
	\end{dmath}
	\begin{dmath}
		H_{\mu \nu}^{(2)}=a_{2}  \Big(2  (g_{\mu \nu} \Box - \nabla_{\mu} \nabla_{\nu}) R + 2 R (R_{\mu \nu} - \frac{1}{4} R g_{\mu \nu})  \Big)\,,
	\end{dmath}
	\begin{dmath}
		H_{\mu \nu}^{(3)}=b_{1}  \Big(2 \nabla_{\mu}{\nabla_{\nu}{\Box{R}}}  - 2 R_{\mu \nu} \Box{R} + \nabla_{\mu}{R} \nabla_{\nu}{R} - (  2 \Box^{2}{R} + \frac{1}{2} \nabla^{\gamma}{R} \nabla_{\gamma}{R}) g_{\mu \nu}  \Big)\,,
	\end{dmath}
	\begin{dmath}
		H_{\mu \nu}^{(4)}=b_{2}  \Big(- \Box^{2}{R_{\mu \nu}}+2 \nabla^{\gamma}{\nabla_{(\mu}{\Box{R_{\nu) \gamma}}}} 
		- 2 R^{\gamma}{}_{(\mu} \Box{R_{\nu) \gamma}}  - 2 R^{\alpha}{}_{(\mu} \nabla^{\beta}{\nabla_{\nu)}{R_{\alpha \beta}}} 
		+ 2 R^{\alpha \beta} \nabla_{\alpha}{\nabla_{(\mu}{R_{\nu) \beta}}} 
		+ \nabla^{\gamma}{R} \nabla_{(\mu}{R_{\nu) \gamma}} + \nabla_{\mu}{R^{\alpha \beta}} \nabla_{\nu}{R_{\alpha \beta}} - 2 \nabla_{(\mu}{R^{\alpha \beta}} \nabla_{\alpha}{R_{\nu) \beta}} -
		(  \nabla^{\alpha}{\nabla^{\beta}{\Box{R_{\alpha \beta}}}}+  \frac{1}{2} \nabla^{\gamma}{R^{\alpha \beta}} \nabla_{\gamma}{R_{\alpha \beta}}) g_{\mu \nu}   \Big) \,,
	\end{dmath}
	\begin{dmath}
		H_{\mu \nu}^{(5)}=c_{1}  \Big (3 R_{\mu \alpha} R^{\alpha \beta} R_{\beta \nu}+ \frac{3}{2} g_{\mu \nu} \nabla_{\alpha} \nabla_{\beta} (R^{\alpha \gamma} R_{\gamma}^{\beta}) +\frac{3}{2} \Box{R^{\gamma}{}_{\mu} R_{\gamma \nu}}  -6 \nabla_{\alpha} \nabla_{(\mu} (R^{\beta}{}_{\nu)} R^{\alpha}{}_{\beta}) -\frac{1}{2} g_{\mu \nu} R^{\alpha}{}_{\beta} R^{\beta \gamma} R_{\gamma \alpha}  \Big)\,,
	\end{dmath}
	\begin{dmath}
		H_{\mu \nu}^{(6)}=c_{2}  \Big((R_{\mu \nu}- \frac{1}{2} g_{\mu \nu} R) R_{\alpha \beta} R^{\alpha \beta} +2 R R^{\gamma}{}_{\mu} R_{\gamma \nu} + g_{\mu \nu} \nabla_{\alpha} \nabla_{\beta} (R^{\alpha \beta} R) + \Box(R R_{\mu \nu}) -2 \nabla_{\gamma} \nabla_{(\mu} (R^{\gamma}{}_{\nu)} R) 
		+ (g_{\mu \nu} \Box - \nabla_{\mu} \nabla_{\nu} ) (R_{\alpha \beta} R^{\alpha \beta})  \Big)\,,
	\end{dmath}
	\begin{dmath}\label{EOM7}
		H_{\mu \nu}^{(7)}=c_{3}  \Big( 3 R_{\mu \nu} R^{2}+3 (g_{\mu \nu} \Box - \nabla_{\mu} \nabla_{\nu}) R^{2}- \frac{1}{2} g_{\mu \nu} R^{3} \Big)\,.
	\end{dmath}
\end{dgroup}

\subsection{The linearized  equations of motion }

Let us  consider a maximally symmetric space in $d$ dimensions as a solution to the equations of motion (\ref{EOM}). The Riemann, Ricci and scalar curvature tensors can be written as
\bea\label{maxim}
R_{\alpha \mu \beta \nu} = \Lambda (g_{\alpha \beta} g_{\mu \nu} - g_{\mu \beta} g_{\alpha \nu} )\,,\quad 
R_{\alpha \beta} =\Lambda (d-1) g_{\alpha \beta}\,,\quad
R=\Lambda d (d-1)\,,
\eea  
where $\L$ is the cosmological constant.
By inserting the above tensors into the equations of motion (\ref{EOM1})-(\ref{EOM7})  we will find that the cosmological parameter $\Lambda_{0}$ is related to the cosmological constant via
\begin{dmath}\label{Lambda0}
	\Lambda_{0}=\frac{1}{2} (d-1) \Lambda \Big ( (d-2) \sigma + (d-4) (d-1)  \Lambda (a_{1}+d a_{2}) 
	+(d-6)(d-1)^2 \Lambda^{2} (c_{1} + d c_{2} + d^2 c_{3})\Big) \,.
\end{dmath}
Now we suppose that the metric fluctuations $h_{\mu \nu}$ are around an $AdS_d$ background $\bar{g}_{\mu \nu}$ which its radius has been fixed by relation (\ref{Lambda0}) and the metric is given by
$g_{\mu \nu}= \bar{g}_{\mu \nu} + \kappa h_{\mu \nu}$.
 If we insert this into the equation of motion (\ref{EOM}) we will find the linearized equation of motion as follow\footnote{We have used the same approach and notation as \cite{Bergshoeff:2012ev}}
\begin{dmath}\label{lineom}
 \mathcal{E}^{L}_{\mu \nu}=\bar\sigma\mathcal{G}_{\mu \nu}(h)+\sigma_{1} \mathcal{G}_{\mu \nu }(\mathcal{G}(h))+\sigma_{2}\mathcal{G}_{\mu \nu}(\mathcal{G}(\mathcal{G}(h)))  \\
+\sigma_{3} (\bar{g}_{\mu \nu} \bar\Box - \bar\nabla_{\mu}\bar\nabla_{\nu} + (d-1) \Lambda \bar{g}_{\mu \nu} )R^{(1)} \\
+\sigma_{4} (\bar{g}_{\mu \nu} \bar\Box - \bar\nabla_{\mu} \bar\nabla_{\nu} + (d-1) \Lambda \bar{g}_{\mu \nu} \bar\Box R^{(1)}\,.
\end{dmath}
In above equation the various constants have been defined as follows
\begin{align}\label{sigmas}
\bar\sigma &=\sigma + (d - 1) \Lambda \Big(2 a_{1} + 2d a_{2} + 3(d- 1)\Big(c_{1} + d(c_{2} + d c_{3})\Big) \Lambda \Big)\,,\nn \\
\sigma_{1}&=-2\Big(a_{1} -\Big(2 b_{2} - (d- 1)(3 c_{1} + d c_{2} )\Big) \Lambda \Big)\,,\qquad \sigma_{2}=- 4b_{2} \,,\nn\\
\sigma_{3}&=\frac{1}{2} \Big(4 a_{2} -(d-4) a_{1} -\Big(b_{2} (d-3)(d-2)^2+(d-1)\big(3 d c_{1}\\
&\,\,\,\,-12(c_{1}+d c_{3})+c_{2}(d(d-4)-8)\big)\Big)\Big)\,,\nn \\  \sigma_{4}&=\frac{1}{2}\Big(-4b_{1}-b_{2} (d (d-5)+8)\Big)\,.\nn
\end{align}
In equation (\ref{lineom}) we have used $\mathcal{G}_{\mu \nu}(h)$ as a linearized expression for the Einstein tensor which define by \cite{Deser:2011xc},
\be\label{ETensor}
G_{\mu \nu}=R_{\mu \nu}-\frac{1}{2} R g_{\mu \nu} +\frac{1}{2} (d-1)(d-2) \Lambda g_{\mu \nu}\,,
\ee
Therefore the  linearized form of the  Einstein tensor is given by
\begin{align}\label{GL}
\mathcal{G}_{\mu \nu}(h)&=R^{(1)}_{\mu \nu}-\frac{1}{2} R^{(1)} \bar{g}_{\mu \nu} - (d-1) \Lambda h_{\mu \nu}\nn\\
&=\bar\nabla_{\alpha} \bar\nabla_{(\mu} h_{\nu)}^{\alpha}-\frac{1}{2} \bar\Box{h_{\mu \nu}}-\frac{1}{2} \bar g_{\mu \nu} (\bar\nabla_{\alpha} \bar\nabla_{\beta} h^{\alpha \beta} - \bar\Box{h})\nn \\
& -\frac{1}{2} \bar\nabla_{\mu}\bar\nabla_{\nu} h+\frac{1}{2} (d-1)(\bar g_{\mu \nu} h -2 h_{\mu \nu} )\Lambda \,,
\end{align}
where we have used the following linearized Ricci and scalar curvature tensors 
\bea\label{linR}
&&R^{(1)}_{\mu \nu}=\frac{1}{2} (\bar\nabla^{\sigma}{\bar\nabla^{\mu}{h_{\nu \sigma}}} + \bar\nabla^{\sigma}{\bar\nabla^{\nu}{h_{\mu \sigma}}} - \bar\Box{h_{\mu \nu}} - \bar\nabla_{\mu}{\bar\nabla_{\nu}{h}})\,, \nn\\
&& R^{(1)}= - \bar\Box{h} + \bar\nabla^{\sigma}{\bar\nabla^{\mu}{h_{\sigma \mu}}} - (d-1) \Lambda  h\,.
\eea
\subsection{Massless and massive graviton modes}
By multiplying the equation (\ref{lineom}) with $\bar{g}^{\m\n}$ one can find the trace of linearized equation of motion in terms of covariant derivatives of the linearized scalar 
 curvature tensor in (\ref{linR})
\be\label{treom}
z_{1} R^{(1)}+ z_{2} \bar\Box R^{(1)}+ z_{3} \bar\Box^2 R^{(1)}=0\,,
\ee
where
\begin{align}
z_{1}& = -\frac{1}{2} (d-2) \bar \sigma-\frac{1}{4} (d-1)(d-2)^2 \Lambda \sigma_{1} -\frac{1}{8} (d-1)^2(d-2)^3 \Lambda \sigma_{2}+
d (d-1) \Lambda \sigma_{3}     \,, \nn \\
z_{2} & =- \frac{1}{4} (d-2)^2 \sigma_{1} -\frac{1}{4} (d-1)(d-2)^3 \Lambda \sigma_{2} + (d-1) \sigma_{3} + d (d-1) \Lambda \sigma_{4}  \,,\nn\\
z_{3}& = - \frac{1}{8} (d-2)^3 \sigma_{2} + (d-1) \sigma_{4} \,. 
\end{align}
As indicated in  \cite{Deser:2011xc}, in order to avoid the propagating scalar degrees of freedom in $AdS_d$ background we will restrict ourselves to the parameters that satisfy the relation  $z_3=z_2=0$ or
\bea\label{a2b1}
&& a_{2}=\frac{1}{{4(d-1)}}\big( \big(b_{2} (d - 2)^2 - (d - 1)\big(3d(c_{1}+ 4c_{3}(d - 1)) 
\nn \\
&&+ c_{2}(d^2 + 8d - 8)\big)\big) \Lambda-d a_{1}\big)\,, \qquad  b_{1}=-\frac{d}{4(d-1)} b_{2}\,.
\eea

With these conditions, the D'Alembertian operator will be removed from equation (\ref{treom}) and therefore the trace of linearized equation of motion reduces to a simpler form, $z_{1} R^{(1)}=0$.
We also assume that $z_{1}\neq0$, therefore $R^{(1)}$ must be vanished.
As noted in  \cite{Gullu:2009vy} we may choose the gauge condition $ \bar\nabla^{\mu}{h_{\mu \nu}}=\bar\nabla_{\nu}{h} $, which from (\ref{linR}) one leads to $ R^{(1)}=- (d-1) \Lambda h $ and therefore  
one can set  $h=0$. Consequently the gauge condition for $h_{\mu \nu}$ would be the transverse and traceless gauge  $ \bar\nabla^{\mu}{h_{\mu \nu}}=h=0$.
The linearized Ricci and Einstein tensors in this  transverse-traceless gauge become
\bea
R^{(1)}_{\mu \nu}=d \Lambda h_{\mu \nu} -\frac{1}{2} \bar\Box{h_{\mu \nu}}\,,\qquad
\mathcal{ G}_{\mu \nu}=\Lambda h_{\mu \nu} -\frac{1}{2} \bar\Box{h_{\mu \nu}}\,,
\eea
and the linearized equation of motion (\ref{lineom})  simplifies to
\begin{align}\label{glineom}
 \mathcal{E}_{\mu \nu}^{L}&= -\frac{\sigma_{2}}{8}  \bar \Box^{3} h_{\mu \nu}+ \frac{1}{4} (  \sigma_{1} + 3 \Lambda \sigma_{2})   \bar \Box^{2} h_{\mu \nu} -\frac{1}{2} ( \bar\sigma +2\Lambda \sigma_{1} +3 \Lambda^{2} \sigma_{2} )  \bar \Box h_{\mu \nu}\nn \\
&+\Lambda ( \bar \sigma + \Lambda \sigma_{1} + \Lambda^{2} \sigma_{2}) h_{\m\n} \,.
\end{align}
As we see, this equation depends  on three parameters $\bar{\s}, \s_1$ and $\s_2$ where we have defined in equation (\ref{sigmas}).
The linearized equation of motion (\ref{glineom}) now can be rewritten as   
\bea\label{3Box}
&&-\frac{\sigma_{2}}{8} (\bar\Box - 2 \Lambda ) (\bar\Box -2 \Lambda - M_{+}^{2} ) (\bar\Box - 2 \Lambda - M_{-}^{2}) h_{\mu \nu}=0\,,
\eea
so that the massless and massive modes satisfy the following Klein-Gordon equations in $AdS_d$ background
\bea\label{Mgraviton}
&& (\bar\Box - A) h^{0}_{\mu \nu}=0 \,,\nn\\&& (\bar\Box - 2 \Lambda - M_{+}^{2}) h^{M_{+}}_{\mu \nu}=0\,,\\&& (\bar\Box -2 \Lambda - M_{-}^{2}) h^{M_{-}}_{\mu \nu}=0\,,\nn
\eea
where the values of masses can be read as
\be\label{mpm}
M_{\pm}^2=\frac{\sigma_{1}\pm \sqrt{\sigma_{1}^2-4\bar\sigma \sigma_{2}}}{\sigma_{2}}\,.
\ee
As we see, the parameter space which defined by parameters $\{\s, a_1, b_2, c_1, c_2, c_3\}$, now can be considered as a space with parameters $\{\bar\s, \s_1,\s_2\}$ when we study the mass of graviton modes. In order to have a free tachyon condition we must restrict ourselves to $M^2_{\pm}\geq 0$ together with $\s_1^2\geq 4\bar{\s} \s_2$. We have summarized the analysis of these conditions in table 1. This table shows that the only allowed regions are those with all values of $\{\bar\s, \s_1,\s_2\}$ positive or all negative.

	\begin{table}
		\centering
\begin{tabular}{|c||c|c|c|}
	\hline 
$\s_1^2\geq 4\bar{\s} \s_2$	& $\s_1$ & $\s_2$  & $\bar{\s}$    \\ 
	\hline  
$(M^2_{+}>0, M^2_{-}>0)^*$& $+$ & $+$ & $+$ \\ 
	\hline 
$(M^2_{+}>0, M^2_{-}>0)^*$& $-$ & $-$ & $-$ \\
	\hline
$(M^2_{+}<0, M^2_{-}<0)$& $+$ & $-$ & $-$ \\ 
\hline 
$(M^2_{+}<0, M^2_{-}<0)$& $-$ & $+$ & $+$ \\
\hline
\end{tabular}
\quad
\begin{tabular}{|c||c|c|c|}
	\hline 
	$\s_1^2\geq 4\bar{\s} \s_2$	& $\s_1$ & $\s_2$  & $\bar{\s}$    \\ 
	\hline  
	$(M^2_{+}>0, M^2_{-}<0)$& $+$ & $+$ & $-$ \\ 
	\hline 
	$(M^2_{+}>0, M^2_{-}<0)$& $-$ & $+$ & $-$ \\
	\hline
	$(M^2_{+}<0, M^2_{-}>0)$& $+$ & $-$ & $+$ \\ 
	\hline 
	$(M^2_{+}<0, M^2_{-}>0)$& $-$ & $-$ & $+$ \\
	\hline
\end{tabular}
\centering
\caption{Tachyon free conditions in parameter space.}
\label{tab:table1}
\end{table}

There are special subspaces in this three-parameter space:

$\bullet{}$ At $\sigma_{1} = \bar\sigma= 0$ and for $\sigma_{2} \neq 0$ in this parameter space, $M_{\pm}^{2}=0$. This  corresponds to a tricritical point  where two massive modes degenerate into the massless one. At this point the massive gravitons are replaced by new solutions, called ``$\log$" and ``$\log^2$" ghost modes, for example see \cite{Bergshoeff:2012ev}. The linearized equation of motion at this point has a simple form of an equation of motion for a spin 2 version of the 3-rank scalar field, i.e. $\mathcal{G}_{\mu \nu}(\mathcal{G} (\mathcal{G}(h)))=0$.

$\bullet{}$ One can find another critical subspace in the parameter space as ($\sigma_{2} \neq 0$, $\bar\sigma \neq 0$) at
$\sigma_{1} ^{2} = 4 \bar \sigma \sigma_{2}$ which in this case, two massive gravitons degenerate into each other, i.e. $M_{+}^{2} = M_{-}^{2}=\s_1/\s_2$.

$\bullet{}$ Moreover we have another critical subspace which is defined by ($\sigma_{2} \neq 0$, $\sigma_{1} \neq 0$ ) at $\bar \sigma = 0$, 
where one of the massive modes degenerates into the massless mode, $M_{-}=0$ and $M_{+}=2\s_1/\s_2$. In this critical line, the degenerated graviton is a logarithmic ghost mode.

$\bullet{}$ In a special situation, when $\s_2=0$ the linearized equation of motion reduces to
\begin{align}
\mathcal{E}_{\mu \nu}^{L}&=\frac{1}{4} \sigma_{1} \bar \Box^{2} h_{\mu \nu} -\frac{1}{2} ( \bar\sigma +2\Lambda \sigma_{1})  \bar \Box h_{\mu \nu}
+\Lambda ( \bar \sigma + \Lambda \sigma_{1}) h_{\m\n} \,,\nn\\
&=\frac{\sigma_{1}}{4} (\bar\Box - 2 \Lambda ) (\bar\Box -2 \Lambda - \hat{M}^{2} ) h_{\mu \nu}=0\,,
\end{align}
where we have just one massive mode with $\hat{M}^2=2\bar{\s}/\s_1$. The stability holds here when both $\s$ and $\s_1$ parameters are positive or negative.

As we mentioned in introduction, there will be a unitary subspace if and only if the ghost-like logarithmic modes at the critical points are truncated by imposing certain boundary conditions  \cite{Bergshoeff:2012sc,Bergshoeff:2012ev,Nutma:2012ss,Kleinschmidt:2012rs,Lu:2013hx}.
But it should be noted that the unitary truncation method is valid only in free theories at the linear level \cite{Apolo:2012vv}.
\subsection{Conserved charges}
In order to obtain the conserved charges corresponding to the symmetries of the theory,  following \cite{Abbott:1981ff}, \cite{Deser:2002rt} and \cite{Deser:2002jk},  we may suppose a Killing vector $\xi_{\nu}$ and use the linearized equation of motion to write $\xi_{\nu} \mathcal{E}^{\mu \nu}_{L} $ as a surface integral. We use this method to find  the mass of asymptotically Schwarzschild-AdS  black holes in Ricci cubic gravity.

In the Abbott-Deser method \cite{Abbott:1981ff} the linearized equation of motion ${\cal{E}}^{L}_{\mu \nu}$ is considered as an effective energy-momentum tensor. This allows us to compute the conserved charges $Q^{\mu}$
as follow
\be\label{ccharge}
Q^{\mu}(\xi)=\int_{\Sigma}{d^{d-1}}x \sqrt{-\bar{g}} \xi_{\nu} \mathcal{E}_{L}^{\mu \nu} \,,
\ee
where $\Sigma$ is a spatial $(d-1)$ dimensional hypersurface.
For calculating the conserved charges, one can show that the integrand can be written as a divergence of a two-form i.e. $\xi_{\nu}\mathcal {E}_{L}^{\mu \nu} =\bar \nabla_{\nu} \mathcal{F}^{\mu \nu}$.
Therefore the integral in (\ref{ccharge})  reduces to a surface integral at the spatial infinity
\be
Q^{\mu}(\xi)=\int_{\partial \Sigma}{dS_{\alpha} \mathcal{F}^{\mu \alpha}}\,,
\ee
where $\partial \Sigma$ is the $(d-2)$ dimensional boundary of $\Sigma$. The conserved charge associated to the RCG can be found by this method from the linearized equation of motion (\ref{lineom}) as
\begin{align}\label{ccharge1}
Q^{\mu}( \xi)&=\frac{1}{4 \Omega_{d-2} G_{d}} \int_{\Sigma}{d^{d-1}}x \sqrt{- \bar{g}} \Big( \bar \sigma  \xi_{\nu} \mathcal{G}^{\mu \nu}(h) + \sigma_{1} \xi_{\nu}\mathcal{G}^{\mu \nu}(\mathcal{G}(h)) \nn\\
&+ \sigma_{2} \xi_{\nu} \mathcal{G}^{\mu \nu}(\mathcal{G}(\mathcal{G}(h))) + \sigma_{3} \xi_{\nu}(\bar{g}^{\mu \nu} \bar\Box - \bar\nabla^{\mu} \bar\nabla^{\nu} + (d-1) \Lambda \bar{g}^{\mu \nu}) R^{(1)}\nn\\
&+ \sigma_{4} \xi_{\nu}(\bar{g}^{\mu \nu} \bar\Box - \bar\nabla^{\mu} \bar\nabla^{\nu} + (d-1) \Lambda \bar{g}^{\mu \nu})  \bar\Box{R^{(1)}} \Big)\,,
\end{align}
where we have found this result by generalizing  the approach of \cite{Bergshoeff:2012ev} to the $d$ dimensional space-time. The overall factor is chosen for future proposes in computing the mass of black hole solutions. The equation (\ref{ccharge1}) is written to the form of  $\bar \nabla_{\nu} \mathcal{F}^{\mu \nu}$ through the following relations
\begin{dgroup}\label{Qchargea}
\begin{dmath}
\xi_{\nu}\mathcal{G}^{\mu \nu}(h)=\bar \nabla_{\rho }\Big (\xi_{\nu} \bar \nabla^{[\mu}{h^{\rho] \nu}}+ \xi^{[\mu} \bar\nabla^{\rho]}{h}+ h^{\nu[ \mu} \bar\nabla^{\rho]}{\xi_{\nu}}-\xi^{[\mu}  \bar\nabla_{\nu}{h^{\rho] \nu}}+ \frac{1}{2} h\bar\nabla^{\mu}{\xi^{\rho}}  \Big) \,,
\end{dmath}
\begin{dmath}
\xi_{\nu}\mathcal{G}^{\mu \nu}(\mathcal{G}(h))=\bar \nabla_{\rho }\Big (\xi_{\nu} \bar \nabla^{[\mu}{\mathcal{G}^{\rho] \nu}}(h)+ \xi^{[\mu} \bar\nabla^{\rho]}{\mathcal{G}(h)}+ \mathcal{G}^{\nu[ \mu}(h) \bar\nabla^{\rho]}{\xi_{\nu}}
\!-\!\xi^{[\mu} \bar\nabla_{\nu}{\mathcal{G}^{\rho] \nu}}(h)+ \frac{1}{2} \mathcal{G}(h)\bar\nabla^{\mu}{\xi^{\rho}} \Big),
\end{dmath}
\begin{dmath}
\xi_{\nu}\mathcal{G}^{\mu \nu}(\mathcal{G}(\mathcal{G}(h)))=\bar \nabla_{\rho }\Big (\xi_{\nu} \bar \nabla^{[\mu}{\mathcal{G}^{\rho] \nu}}(\mathcal{G}(h))+ \xi^{[\mu} \bar\nabla^{\rho]}{\mathcal{G}(\mathcal{G}(h))}+ \mathcal{G}^{\nu[ \mu}(\mathcal{G}(h)) \bar\nabla^{\rho]}{\xi_{\nu}}
-\xi^{[\mu}  \bar\nabla_{\nu}{\mathcal{G}^{\rho] \nu}} \mathcal{G}(h)
+ \frac{1}{2} \mathcal{G}(\mathcal{G}(h))\bar\nabla^{\mu}{\xi^{\rho}} \Big), 
\end{dmath}
\begin{dmath}
\xi_{\nu}(\bar{g}^{\mu \nu} \bar\Box - \bar\nabla^{\mu} \bar\nabla^{\nu} + (d-1) \Lambda \bar{g}^{\mu \nu}) R^{(1)}= -\frac{4}{d-2} \bar\nabla_{\rho}\Big(\xi^{[\mu} \bar\nabla^{\rho]}{\mathcal{G}(h)} +\frac{1}{2} \mathcal{G}(h) \bar\nabla^{\mu} \xi^{\rho}\Big)\,,
\end{dmath}
\begin{dmath*}
 \xi_{\nu}(\bar{g}^{\mu \nu} \bar\Box - \bar\nabla^{\mu} \bar\nabla^{\nu} + (d-1) \Lambda \bar{g}^{\mu \nu})\bar \Box{R^{(1)}}= -\frac{4}{d-2} \bar\nabla_{\rho}\Big(\xi^{[\mu} \bar\nabla^{\rho]}\bar\Box{\mathcal{G}(h)} +\frac{1}{2} \bar\Box \mathcal{G}(h) \bar\nabla^{\mu} \xi^{\rho}\Big)\,,
\end{dmath*}
\begin{dmath}
\,	
\end{dmath}
\end{dgroup}
where $\mathcal{G}_{\m\n}(h)$ was introduced in equation (\ref{GL}).
Finally the conserved quantities can be found from the following relation
\bea\label{Qchargeb}
Q^{\mu}( \xi)\!\!\!\!\! \!\!\!&&=\frac{1}{4 \Omega_{d-2} G_{d}} \int_{\Sigma}{d^{d-1}}x \sqrt{- \bar{g}} \Big( \bar \sigma  \xi_{\nu} \mathcal{G}^{\mu \nu}(h) + \sigma_{1} \xi_{\nu}\mathcal{G}^{\mu \nu}(\mathcal{G}(h)) 
+ \sigma_{2} \xi_{\nu} \mathcal{G}^{\mu \nu}(\mathcal{G}(\mathcal{G}(h)))\Big) \nn \\ 
&&-\frac{1}{4 \Omega_{d-2} G_{d}} \int_{\partial \Sigma}{dS_{\rho}} \sqrt{- \bar{g}} \Big (
\frac{4\sigma_{3} }{d-2} \big (\xi^{[\mu} \bar\nabla^{\rho]}{\mathcal{G}(h)} +\frac{1}{2} \mathcal{G}(h) \bar\nabla^{\mu} \xi^{\rho}\big) \nn \\ 
&&+ \frac{4\sigma_{4}}{d-2} \big(\xi^{[\mu} \bar\nabla^{\rho]}\bar\Box{\mathcal{G}(h)} +\frac{1}{2} \bar\Box \mathcal{G}(h) \bar\nabla^{\mu} \xi^{\rho}\big) \Big)\,. 
\eea
We will use this relation to compute the mass of black holes  in $AdS_d$ space-time.

\section{Auxiliary field formalism}

In this section we are going to rewrite the Ricci cubic action to the form of Fierz-Pauli massive action for spin-two fields. Writing in this form we will be able again to calculate the mass of graviton modes. We will do this by employing the auxiliary field formalism. 

For this purpose we need to reformulate the six derivatives action (\ref{action}) with using the auxiliary fields which produce an action with just  second order  derivative terms.
To do this, we need to introduce two rank-two auxiliary fields $(f_{\m\n} , \l_{\m\n})$ \cite{Hohm:2010jc}, \cite{Bergshoeff:2012ev} and \cite{Bergshoeff:2011ri}. Let us start from the following action\footnote{Note that our choice for those terms in \ref{auxantz} with covariant derivative of $\l_{\m\n}$, differs from  the choice of \cite{Bergshoeff:2012ev}. We have considered all possible terms and we do not need to add extra boundary terms like those which appeared in \cite{Bergshoeff:2012ev}. }

\begin{dmath}\label{auxantz}
S=\frac{1}{\k^2}\int d^d x \sqrt{-g}\Big(  
\sigma R-2 \Lambda_0 + \c_1 f^{\alpha\beta} R_{\alpha\beta} + \c_2 fR+ \c_3 f^{\alpha\beta} \lambda_{\alpha\beta}  + \c_4 f\lambda+ \c_5 \lambda_{\alpha \beta} \lambda^{\alpha \beta} + \c_6 \lambda^2  
  + \c_7 \nabla^{\m}\lambda^{\alpha \beta} \nabla_{\m} \lambda_{\alpha \beta} 
  + \c_8 \nabla^{\m}\lambda \nabla_{\m}\lambda  
  + \c_9 \nabla_{\alpha}\lambda^{\alpha \beta} \nabla^{\mu}\lambda_{\mu \beta}
  + \c_{10} \nabla^{\mu} \lambda\nabla^{\nu}\lambda_{\mu \nu}
  +\c_{11} \nabla_\b \l_{\a\m} \nabla^\m \l^{\a\b}
  + \c_{12} \lambda^3 
  + \c_{13} \lambda \lambda_{\mu \nu} \lambda^{\mu \nu}
  + \c_{14} \lambda_{\alpha \beta} \lambda^{\beta \mu} \lambda_{\mu}{}^{\alpha} \Big)\,,
\end{dmath}
where $f$ and  $\lambda$ are traces of the auxiliary fields. 
We can find the unknown coefficients by computing the equations of motion for auxiliary fields in $d$ dimensions as follows
\begin{dgroup}
\begin{dmath}\label{lfeom1}
\lambda_{\mu \nu} = \frac{1}{d-2 } \Big(R_{\mu \nu} -\frac{1}{2(d-1)} R g_{\mu \nu}\Big)\,, 
\end{dmath}
\begin{dmath}\label{lfeom2}
f_{\mu \nu} = \frac{1}{(d-2)(d-1)} \Big ((d-1) \Big( 2 (\c_5 + \c_{13} \lambda) \lambda_{\m\n} + 3 \c_{14}  {\lambda_\m}^{\delta} \lambda_{\n\delta}
+  2 \c_7 \Box \lambda_{\m\n}+ 2\c_9 \nabla_{(\m}\nabla_{\delta}{\lambda_{\n)}}^{\delta}+\c_{10} \nabla_{\m}\nabla_{\n}\lambda 
+ 2\c_{11} \nabla_{\delta}\nabla_{(\m}{\lambda_{\n)}}^{\delta}
\Big) 
- g_{\m\n} \Big(\big(2 (\c_5 + \c_6) + (3 \c_{12} + 2 \c_{13}) \lambda\big)\lambda  +( \c_{13} + 3 \c_{14}) \lambda_{\a\b} \lambda^{\a\b}
+  \big(\c_{10} + 2 (\c_7 + \c_8)\big) \Box\lambda- (\c_{10} + 2 (\c_9+\c_{11})) \nabla_{\a}\nabla_{\b}\lambda^{\a\b} \Big)\Big)\,,
\end{dmath}
\end{dgroup}
where we have fixed 
\be
\c_1=1\,,\qquad \c_2=-\frac12\,,\qquad \c_3=-(d-2)\,,\qquad \c_4=(d-2)\,,
\ee
by using the freedom in scaling of the fields and demanding that the equation of motion from variation of $f_{\a\b}$, gives the value of auxiliary field $\l_{\m\n}$  equal to  the Schouten tensor in $d$ dimensions.
By inserting the above results into the Lagrangian (\ref{auxantz}) and comparing with the Lagrangian in the original action (\ref{action}) one finds the following values 
\begin{align}\label{chis1}
&\c_5=a_1 (d-2)^2\,,\qquad \c_6=4 a_2 (d-1)^2+a_1 (3 d-4) \,, \nn \\
&\c_7=b_2 (d-2)^2\,,\qquad \c_9=-\c_{10}-\c_8+b_2 (3 d-4)+4 b_1 (d-1)^2\,, \nn \\
&\c_{11}=0\,,\qquad \c_{12}=c_1 (4d-6)+2 (d-1) (4 c_3 (d-1)^2+c_2 (3 d-4))\,, \nn \\
&\c_{13}=(3 c_1+2 c_2 (d-1)) (d-2)^2\,, \qquad \c_{14}=c_1 (d-2)^3\,,
\end{align}
where as we see, all coefficients have been fixed except $\c_8$ and $\c_{10}$. In next section we will show that we are able to fix these remaining coefficients too.
\subsection{Graviton mass spectrum }
Now we can expand the new action (\ref{auxantz}) around the $AdS_{d}$ maximally space up to the second order of field perturbations.
The perturbation of auxiliary fields around their background values can be defined through a linear combination of two fluctuating fields ${k_1}_{\m\n}$ and ${k_2}_{\m\n}$ together with  the background metric perturbation $h_{\m\n}$, i.e.
\begin{align}\label{k1k2}
\lambda_{\mu \nu} &= \frac{\Lambda}{2} (\bar{g}_{\mu \nu} + \kappa h_{\mu \nu}) + \kappa k_{1 \mu \nu} \,,\qquad 
f_{\mu \nu} = \zeta \Lambda  (\bar{g}_{\mu \nu} + \kappa h_{\mu \nu}) + \kappa k_{2 \mu \nu} \,, \nn \\
\zeta&=-\frac{2\L (d-1)}{d-2}(2 (a_1 +  d a_2) + 
3 (d-1) (c_1 +d(c_2 + d c_3)) \L) \,,
\end{align}
where the coefficients are chosen so that the background values satisfy equations (\ref{lfeom1}, \ref{lfeom2}).
By expanding (\ref{auxantz}) around these background fields up to the second order of perturbations and by substituting the following expressions for Ricci tensor 
\bea
&&R_{\mu \nu}= R^{(0)}_{\mu \nu} + \kappa R^{(1)}_{\mu \nu}+\kappa^{2} R^{(2)}_{\mu \nu}\,,\nn \\
&&R^{(0)}_{\mu \nu} =\Lambda (d-1) \bar g_{\mu \nu}\,,
\qquad R^{(1)}_{\mu \nu} =  - \frac{1}{2} (  \Box{h_{\mu \nu}}+ \nabla_{\nu}{\nabla_{\mu}{h}} -  \nabla^{\alpha}{\nabla_{\mu}{h_{\nu \alpha}}} -  \nabla^{\alpha}{\nabla_{\nu}{h_{\mu \alpha}}})\,,\nn \\
&&
R^{(2)}_{\mu \nu} =\frac{1}{4} \Big( \nabla_{\m}h^{\a\b} \nabla_{\n}h_{\a\b}+\nabla^{\a}h_{\m\n} (2\nabla_{\b}h_{\a}{}^{\b}- \nabla_{\a}h ) + 2 (\nabla_{\b}h_{\n\a}  -  \nabla_{\a}h_{\n\b}) \nabla^{\b}h_{\m}{}^{\a} \nonumber \\ 
&& + 2(\nabla_{\a}h  - 2 \nabla_{\b}h_{\a}{}^{\b}) \nabla_{(\m}h_{\n)}{}^{\a} + 2 h^{\a\b} (\nabla_{\b}\nabla_{\a}h_{\m\n} -  2\nabla_{\b}\nabla_{(\m}h_{\n)\a}   + \nabla_{\m}\nabla_{\n}h_{\a\b})\Big)\,,
\eea
we will obtain the following Lagrangian 
\begin{align}\label{L2}
\mathcal{L}^{(2)}&= -\frac{1}{2} \bar\sigma h^{\mu \nu} \mathcal{G}_{\mu \nu}(h) + \xi_1 k_{1}^{\mu \nu} \mathcal{G}_{\mu \nu}(k_{1}) +  \xi_2 k_{2}^{\mu \nu} \mathcal{G}_{\mu \nu}(h)\nn \\ 
&+ \xi_3 (k_{1}^{\mu \nu} k_{1 \mu \nu} - k_{1} k_{1} )+\xi_4 (k_{1}^{\mu \nu} k_{2 \mu \nu} - k_{1} k_{2})\,,
\end{align}
where $\mathcal{G}_{\mu \nu}(h)$ is the linearized Einstein tensor  (\ref{GL}) or equivalently
\begin{align}
\mathcal{G}_{\mu \nu}(h)= -\frac{1}{2} ( \Box{h_{\mu \nu}} + \nabla_{\nu}{\nabla_{\mu}{h}} -2 \nabla_{(\mu}\nabla^{\rho}{h_{\rho \nu)}} -2 \Lambda h_{\mu \nu} -  (d-3)  \Lambda \eta_{\mu \nu} h)\,.
\end{align}
To find $ \mathcal{G}_{\mu \nu}(k_{1})$ one needs to replace $h_{\m\n}$ with ${k_1}_{\m\n}$ in above equation. 
The coefficients in  Lagrangian (\ref{L2}) are given by
\begin{align}
\bar\sigma &=  \sigma -\frac{1}{2} (d-2 )^2 \Lambda( a_{1}-d  \Lambda b_{2} ) -\frac{1}{2} (d-1) \Big( 3 ( 2 -2 d + d^{2} ) c_{1} \nn\\ 
&+ d (-2 + 2 d + d^{2}) c_{2} + 6 d^2 (d-1) c_{3}\Big)\Lambda^{2} \,,\\
\xi_1&=2 b_{2}(d-2)^2\,,\quad  \xi_2=1\,,\quad \xi_4=-d+2\,, \nn \\ 
\xi_3&=(d-2)^2(a_{1} + (d-1) ( 3c_{1}  + dc_{2}) \Lambda  -2 \Lambda b_{2}) \,,\nn 
\end{align}
where for computing these coefficients we have used the relation between cosmological constant $\Lambda$ and the cosmological parameter $\Lambda_{0}$  in equation  (\ref{Lambda0}).  We have also used the constraints in (\ref{a2b1}).
In order to write the Lagrangian as the specific form in equation (\ref{L2}),  we need to fix the remaining unfixed coefficients as 
\be \label{chis2}
\chi_8=-b_2 (d-2)^2\,,\qquad \chi_{10}=2b_2 (d-2)^2\,.
\ee
We can go further and write the Lagrangian (\ref{L2}) as a diagonalized form by the following field redefinitions
\begin{align}\label{k1pk2p}
h_{\mu \nu}&=h'_{\mu \nu}+\frac{2 b_2 (d-2) M^{2}_{\pm}}{\bar\sigma} k'_{1 \mu \nu} +\frac{1}{ \bar \sigma}  k'_{2 \mu \nu}\,,\nn\\
k_{1 \mu \nu}&=k'_{1 \mu \nu}- \frac{M^{2}_{\pm}}{2 (d-2) \bar\sigma}k'_{2 \mu \nu}\,,\\
k_{2 \mu \nu}&=k'_{2 \mu \nu}+2 b_2 (d-2) M^{2}_{\pm}k'_{1 \mu \nu}\,,\nn
\end{align}
where immediately give rise to
\begin{align}
\mathcal{G}_{\mu \nu}(h)&=\mathcal{G}_{\mu \nu}(h')+\frac{2 b_2 (d-2) M^{2}_{\pm}}{\bar\sigma }\mathcal{G}_{\mu \nu}(k'_1)+\frac{1}{ \bar \sigma}  
\mathcal{G}_{\mu \nu}(k'_2)\,,\nn\\
\mathcal{G}_{\mu \nu}(k_1)&=\mathcal{G}_{\mu \nu}(k'_1)- \frac{M^{2}_{\pm}}{2 (d-2) \bar\sigma} \mathcal{G}_{\mu \nu}(k'_2)\,.
\end{align}
Doing these, we find a linear combination of massive Fierz-Pauli Lagrangians which contains  a massless spin-2 field  $h'_{\mu \nu}$  and two massive spin-2 fields $k'_{1 \mu \nu}$ and $k'_{2 \mu \nu}$  with  $M^{2}_{\pm}$ mass squares respectively
\begin{align}\label{L2D}
\mathcal{L}^{(2)}=&- \frac{1}{2} \bar \sigma {h'}^{\mu \nu} \mathcal{G}_{\mu \nu}(h') 
+\frac{4 (d-2)^{2} b_2}{\bar \sigma } (\bar \sigma + b_2  M^{4}_{\pm}) ( \frac{1}{2} {k'}^{\mu \nu}_{1} \mathcal{G}_{\mu \nu}({k'}_{1}) -\frac{1}{4} M^{2}_{\mp} ({k'}^{\mu \nu}_{1} {k'}_{1 \mu \nu} - {k'}^{2}_{1})) \nn \\
&+\frac{1}{ \bar \sigma^{2}} ( \bar \sigma + b_2  M^{4}_{\pm})  ( \frac{1}{2} {k'}^{\mu \nu}_{2} \mathcal{G}_{\mu \nu}({k'}_{2}) -\frac{1}{4} M^{2}_{\pm} ({k'}^{\mu \nu}_{2} {k'}_{2 \mu \nu} - {k'}^{2}_{2}))\,.
\end{align}
The achieved values of these masses confirm  exactly the values of mass where we have found from  equation (\ref{mpm})  from  linearizing the equation of motion . In order to have a ghost-free theory we need all kinetic terms to have the same sign. As we see from (\ref{L2D}) for $\bar{\s}\neq0$ this is impossible and we always have a rank two ghost field. This is a general property for higher derivative gravity theories and have been reported in different papers for example see \cite{Bergshoeff:2011ri}.
 
 The holographic studies of critical gravities show that the dual gauge theories are log CFTs, for example see \cite{Deser:2011xc} and \cite{Bergshoeff:2011ri}. For RCG we have found the set of these critical points at the end of subsection 2.2.

\subsection{Energy of the linear excitations }
Using the linearized form of the Lagrangian in (\ref{L2D}) we are able to compute the energy of graviton modes by constructing the Hamiltonian. 
Let's redefine $h'_{\mu \nu}$, the massless mode, by $\psi^{0}_{\mu \nu}$ and massive modes, $k'_{1 \mu \nu}$ and $k'_{2 \mu \nu}$, by $\psi^{\pm}_{\mu \nu}$  as follows
\bea
h'_{\mu \nu}=\psi^{0}_{\mu \nu}\,,\qquad
k'_{1 \mu \nu}=\frac{\bar\sigma }{2 b_2 (d-2) M^{2}_{\pm}} \psi^{+}_{\mu \nu}\,,\qquad
k'_{2 \mu \nu}=\bar \sigma \psi^{-}_{\mu \nu}\,,
\eea
and calculate the Hamiltonian by Ostrogradsky's formalism. We remind that the fields are fixed in the transverse and traceless gauge.
The Hamiltonian is given by
 \begin{align}
H
=\frac{1}{2 \kappa^{2}}\int{d^{d-1}}x \sqrt{- \bar g} \Big[&- \bar \sigma  \dot {h'}_{\mu \nu} \bar \nabla^{0}{{h'}^{\mu \nu} } 
 +\frac{4 b_2 (d-2)^{2} }{\bar\sigma m^{4}} (\bar \sigma 
 + \frac{b_2 M^{4}_{\pm}}{m^{4}}) \dot k'_{1\mu \nu} \bar \nabla^{0}{{k'}^{\mu \nu}_{1} } \nn \\
& + \frac{1}{\bar \sigma^{2}} (\bar \sigma + \frac{b_2 M^{4}_{\pm}}{m^4})  \dot k'_{2\mu \nu} \bar \nabla^{0}{{k'}^{\mu \nu}_{2} }-\mathcal{L}^{(2)}\Big] \,.
 \end{align}
 Therefore the on-shell energies of the linearized modes are
  \begin{dgroup}
 \begin{dmath}\label{Egraviton0}
 E^{0}=- \frac{\bar \sigma}{2 \kappa^{2}} \int{d^{d-1}}x \sqrt{- \bar{g}} \dot \psi^{0}_{\mu \nu} \bar \nabla^{0}{\psi^{0 \mu \nu} } \,, \label{E0}
 \end{dmath}
\begin{dmath}\label{Egraviton1}
 E^{M_{\pm}}= \frac{1}{2 \kappa^{2}} (\bar \sigma + \frac{b_2 M^{4}_{\pm}}{m^4}) \int{d^{d-1}}x \sqrt{- \bar{g}} \dot \psi^{\pm}_{\mu \nu} \bar \nabla^{0}{\psi^{\pm \mu \nu} }\,.\label{EPM}
\end{dmath} 
\end{dgroup}
 To have ghost-free modes, the energy of massless and massive gravitons should have the same sign in equations (\ref{E0}) and (\ref{EPM}). This is  equivalent to demand that all kinetic terms
 in the linear action (\ref{L2D}) must have the same sign. As we told before, in general we have ghost modes in this theory except at the critical points.
The results here has been observed already for $d=3$ in \cite{Bergshoeff:2012ev}.

\section{The boundary stress tensor}
We showed that the generic Ricci cubic curvature theory in arbitrary $d$ dimensions admits a reformulation by using two auxiliary fields in a two-derivatives action . Variation of this action (\ref{auxantz}), produces the following boundary terms
\begin{dgroup}
\begin{dmath*}
\delta{\mathcal{S}^{(b)}}= \frac{1}{\k^2} \int d^{d-1} x \sqrt{-\g}\Big(\mathcal{B}^{1}_{\a}{}^{\b} \d\l^{\a}{}_{\b}+ \mathcal{B}^{2}_{\a\b} \d g^{\a\b}+ \mathcal{B}^{3}_{\a\b\d} \nabla^\a\d g^{\b\d}\Big)\,,
\end{dmath*}
\begin{dmath}
\mathcal{B}^{1}_{\a}{}^{\b} =n^\m \Big(2 \c_{7} \nabla_{\mu}\lambda_{\alpha}{}^{ \beta}\!+\!  g_{\alpha \mu} (\c_{10} \nabla^{\beta}\lambda \!+\! 2 \c_ {9} \nabla^{\delta}\lambda^{\beta}{}_{\delta})
\!+\!  g_{\alpha}{}^{\beta} (\c_{10} \nabla_{\delta}\lambda_{\mu}{}^{\delta} \!+\! 2 \c_{8} \nabla_{\mu}\lambda )\Big), \label{B1} 
\end{dmath}
\begin{dmath}
\mathcal{B}^{2}_{\a\b}=\frac12 n^\m \Big(  \c_1( \nabla_{\mu}f_{\alpha \beta}  - 2 \nabla_{\beta}f_{\m\alpha} +  g_{\alpha \beta} \nabla_{\delta}f_{\mu}{}^{\delta} )+ 2 \c_2 (g_{\alpha \beta} \nabla_{\mu}f  -   g_{\alpha \mu} \nabla_{\beta}f) 
 +4\c_7(\lambda_{\mu}{}^{\delta} \nabla_{\beta}\lambda_{\alpha \delta} -  \lambda_{\alpha}{}^{\delta} \nabla_{\beta}\lambda_{\mu \delta})+2\c_9 (\lambda_{\alpha \mu} \nabla_{\delta}\lambda_{\beta}{}^{\delta} -  
 \lambda_{\alpha \beta} \nabla_{\delta}\lambda_{\mu}{}^{\delta} + ( g_{\alpha \beta}  \lambda_{\mu}{}^{\delta}-  g_{\alpha \mu} \lambda_{\beta}{}^{\delta}) \nabla_{\theta}\lambda_{\delta}{}^{\theta})+\c_{10}(2 \lambda_{\alpha \mu} \nabla_{\beta}\lambda + ( g_{\alpha \beta} \lambda_{\mu}{}^{\delta}-2 g_{\alpha \mu} \lambda_{\beta}{}^{\delta} ) \nabla_{\delta}\lambda -  \lambda_{\alpha \beta} \nabla_{\mu}\lambda)\Big)\,, \label{B2}
\end{dmath}
\begin{dmath}
 \mathcal{B}^{3}_{\a\b\d}=\frac12 n^\m\Big(  
 \c_1 (2 f_{\alpha \beta} g_{\delta \mu}-  f_{\beta \d} g_{\alpha \m} - f_{\alpha \m} g_{\beta \d})+ 2 (\sigma + \c_2 f)( g_{\alpha \d} g_{\beta \m} - g_{\alpha \m} g_{\beta \d})  \Big)\,,\label{B3}
\end{dmath}
\end{dgroup}
where $n^{\mu}$ is a vector normal to the boundary and all coefficients are given in equations (\ref{chis1}) and (\ref{chis2}). In computation of the above expressions we have used ${f^{\m}}_{\n}$ and $\l^\m{}_\n$ as our fundamental fields, see \cite{Hohm:2010jc}.
In order to have a well-defined variational principle we need the generalized Gibbons-Hawking terms\cite{Dyer:2008hb},\cite{Balcerzak:2008bg}.
To do this we employ the method that is introduced in \cite{Hohm:2010jc}.

Let's choose the coordinates $x^{\mu}=(r, x^{i})$ corresponding to a slicing of  $d$-dimensional bulk, by  $(d-1)$-dimensional Lorentzian submanifolds for each value of the radial coordinate $r$. 
We can make an ADM-like split of the metric as
\bea\label{ADM}
ds^{2}=N^{2} dr^{2}+ \gamma_{i j}(dx^{i}+ N^{i} dr) (dx^{j}+ N^{j} dr)\,,
\eea
where $\gamma_{i j}$ defines the boundary metric while $N$ and $N^{i}$ denote the lapse and shift functions respectively. 
Inserting (\ref{ADM}) into (\ref{B3}) one finds
\begin{align}
 \mathcal{B}^{3}_{\a\b\d} \nabla^\a\d g^{\b\d}&=\d K_{ij}\Big(\c_1 f^{ij}+\g^{ij}(2\s+2\c_2 {f^j}_j+(\c_1+2\c_2){f^r}_r)\Big)\nn \\
& -\d \g_{ij}\Big((\s + \c_2  {f^k}_k + (\c_1 + \c_2) {f^r}_r)  K^{ij}-\frac12 n_r \mathcal{D}_i f^{ir} \g^{jk}\Big)\,,
\end{align}
where  $K_{i j}=-\frac12\pa_r \g_{ij}$  is the extrinsic curvature tensor and $\mathcal{D}$ is the covariant derivative with respect to the boundary metric $\gamma_{i j}$. Here we have considered $N=1$ and $N_i=0$ for simplicity but we can always get the generalized results in the final answer similar to the work of \cite{Hohm:2010jc}. By using the above result we can read the Gibbons-Hawking terms  as
\be\label{GH}
\mathcal{S}^{GH}=-\frac{1}{\k^2} \int d^{d-1} x \sqrt{-\g}\Big(\chi_1 f^{ij} K_{ij} + \bigl(2 \sigma + (\chi_1 + 2 \chi_2) f^{r}{}_{r} + 2 \chi_2 f^{i}{}_{i}\bigr) K\Big)\,,
\ee
where $K= K^{i}{}_{i}$.
Now we are able to find the boundary energy-momentum tensor through a variation with respect to the boundary metric $\g_{ij}$
\bea
8 \pi G_{d} T^{i j}=\frac{2}{\sqrt{-\gamma}} \frac{\delta{\mathcal{S}^{tot}}}{\delta{\gamma_{i j}}}\,,\qquad
\delta\mathcal{S}^{tot}=\delta\mathcal{S}^{b}+\delta\mathcal{S}^{GH}\,,
\eea
by using the Gibbons-Hawking terms in equation (\ref{GH}) and boundary terms (\ref{B2}) and  (\ref{B3}) 
\begin{align}\label{BST}
4 \pi G_{d} T^{i j}  &= \sigma \big(K^{ij} - K\gamma^{ij}\big)+\mathcal{T}^{ij}_1+\mathcal{T}^{ij}_2+\mathcal{T}^{ij}_7+\mathcal{T}^{ij}_9
+\mathcal{T}^{ij}_{10}\,, \nn \\ 
\mathcal{T}^{ij}_1  &=   \frac{1}{4} \chi_1 \Big(4 s K^{ij} + 2\mathcal{D}_r f^{ij}  - 4 \mathcal{D}^{(i}h^{j)} -4{f^{(i}}_{k} K^{j)k} +	( 2 \mathcal{D}_r s  - 4 s K + 4  \mathcal{D}_{k}h^{k})\gamma^{ij}\Big)\,, \nn \\
\mathcal{T}^{ij}_2 &=   \chi_2 \Big(s K^{ij} + ( \mathcal{D}_r f- s K + \mathcal{D}_r s) \gamma^{ij} + f (K^{ij} -  K \gamma^{ij})\Big)\,, \nn \\
\mathcal{T}^{ij}_7  &=2\c_7\Big(  - (S^2 K^{ij} + H_{k} H^{k}) K^{ij}  - 3H^{k} H^{(i}  K^{j)}{}_{k} -  K^{(i}{}_{l} \lambda^{j)k} \lambda^{l}{}_{k}\nn \\ 
& -  \lambda^{(ik} \mathcal{D}^{j)}H_{k} + H^{k} \mathcal{D}^{(i}\lambda_{k}{}^{j)} -  H^{(i}\mathcal{D}^{j)}S + S (2K^{(i}{}_{k} \lambda^{j)k}+ \mathcal{D}^{(i}H^{j)})\Big)\,,\nn \\	
\mathcal{T}^{ij}_9 &= \chi_9 \Big(\tfrac{1}{2} \mathcal{D}_r (H^{i} H^{j})  - 2  H^{k} H^{(i} K^{j)}{}_{k} -  H^{i} H^{j} K+  H^{(i} \mathcal{D}_{k}\lambda^{j)k} \nn \\ 
&+ \lambda^{ij} (S K -  \mathcal{D}_rS - K_{kl} \lambda^{kl}  -  \mathcal{D}_{k}H^{k} )  + \gamma^{ij} \big( H^{k} \mathcal{D}_{l}\lambda_{k}{}^{l} +H_{k} \mathcal{D}_rH^{k}
	\hiderel{} \nn \\
	& - 2 H^{k} H^{l} K_{kl} - K( S^2+ H_{k} H^{k})  + S (\mathcal{D}_rS  + K_{kl} \lambda^{kl} + \mathcal{D}_{k}H^{k}) \big)\Big)\,,\nn \\
\mathcal{T}^{ij}_{10} & = \frac12\chi_{10} \Big(-\mathcal{D}_r(S+ \lambda) \lambda^{ij}  +2 H^{(i} \mathcal{D}^{j)}S +2 H^{(i} \mathcal{D}^{j)}\lambda + \gamma^{ij} \big( S\mathcal{D}_r(S+ \lambda) \nn \\ 
&+  H^{k} \mathcal{D}_{k}(S +\lambda)\big)\Big)\,,
\end{align}
where we have used ${\lambda^{i}}_{r}=H^{i}, { \lambda^{r}}_{ r}=S,  
{f^{i}}_{ r}=h^{i}, {f^{r}}_{ r}=s, {f^k}_k=f$ and $ {\l^k}_k=\l$ for simplicity in notation.
We also use   $\mathcal{D}_{r}$ as covariant $r$-derivative and $\mathcal{D}_i$ as covariant derivative with respect to the boundary metric
\be\label{ders}
\mathcal{D}_{r} \lambda^{i j}=\frac{1}{N} \big(\partial_{r} \lambda^{i j}- N^{k} \partial_{k} \lambda^{i j}+ 2 \lambda^{k (j} \partial_{k} N^{i)}) \,, \qquad 
\mathcal{D}_{r} \lambda=\frac{1}{N} \big(\partial_{r} \lambda- N^{j} \partial_{j} \lambda).
\ee
As it has been shown in  \cite{Hohm:2010jc}, in order to take into account the nontrivial lapse and shift functions, it is enough to replace all the above fields with the following combinations of auxiliary fields
\bea\label{hats}
&&\hat \lambda^{i j}=\lambda^{i j}+2 H^{(i} N^{j)}+ s N^{i} N^{j}\,, \qquad  \hat f^{i j}=f^{i j}+2 h^{(i} N^{j)}+ s N^{i} N^{j}\,, \nn\\
&&\hat H^{i}=N(H^{i}+S N^{i})\,,\qquad \hat h^{i}=N (h^{i}+s N^{i})\,,\qquad \hat S=N^{2} S\,,\qquad \hat s=N^{2} s\,.  
\eea

In next section we will use the boundary stress tensor (\ref{BST}) to compute the conserved charges of the RCG for different background space-times. 
To do this, we decompose the boundary geometry in ADM-like form. Consider the boundary coordinates as  $x^i=(t, x^a)$, where $x^a$s belong to the $d-2$ dimensional space-like  hyper-surface $\S$, the metric on the boundary can be written as
\be
\gamma_{i j} d x^{i} d x^{j}=- N^{2}_{0} dt^{2} + \hat{\g}_{ab} (d x^a+ N^a dt)(d x^b+ N^b dt)\,.
\ee
By using a time-like normal vector  $u^{i}$, we can calculate the conserved charges associated to the Killing vector $\xi^{i}$ as
\be\label{Qzeta}
\mathcal{Q}_{\xi}=\int_{\Sigma}d^{d-2}x \sqrt{\hat{\g}} u^{i} T_{i j} \xi^{i}\,.
\ee
For example we can find the mass as follow
\bea\label{NRM}
&&M=\int_{\Sigma}d^{d-2}  x \sqrt{\hat\g} N_{0} T_{i j} u^{i} u^{j}\,.
\eea
We will use this relation to find the mass of different solutions of the RCG such as $SAdS$ and Lifshitz black holes in $d$-dimensions.

In addition to the boundary stress tensor we have two other boundary tensors which achieve by variation with respect to the auxiliary fields $\lambda^{i}_{j}$ and $f^{i}_{j}$
\bea
 8 \pi G_{d}\, {\tau_{1}}^{i}_{j}=\frac{2}{\sqrt{-\gamma}} \frac{\delta{\mathcal{S}^{tot}}}{\delta{\lambda^{i}_{j}}}\,,\qquad
 8 \pi G_{d}\, {\tau_{2}}^{i}_{j}=\frac{2}{\sqrt{-\gamma}} \frac{\delta{\mathcal{S}^{tot}}}{\delta{f^{i}_{j}}}\,.
\eea
By using the action (\ref{GH}) and boundary terms in equation (\ref{B1}) one can find these new boundary tensors as
\begin{align}\label{tau}  
& 8 \pi G_{d}\, {\tau_{1}}^{i}_{j}=2 \chi_{7} \mathcal{D}_{r}{\lambda^{i}_{j}} +
2 \chi_{8}  \mathcal{D}_{r} (\lambda + S) \gamma^{i}_{j} + \chi_{10} (\mathcal{D}_{k}{H^{k}} - K S    +  \mathcal{D}_{r}{S}) \gamma^{i}_{j} \,, \nn\\
& 8 \pi G_{d}\, {\tau_{2}}^{i}_{j}=-2 \chi_{2}  K \gamma_{j}^{i} -\chi_{1} K_{j}^{i} \,.
\end{align}
In holographic dictionary it is well known that the energy-momentum boundary tensor $T^{ij}$  is a holographic response function conjugate to the $h_{ij}$ source. On the other hand the auxiliary field formalism in section 3  and specifically equations (\ref{k1k2}) and (\ref{k1pk2p}) show that a mixed combination of fluctuations of the auxiliary fields is describing the massive graviton modes. So it is probable that a mixture ${\tau_{1}}^{i}_{j}$ and ${\tau_{2}}^{i}_{j}$ play the role of holographic response functions conjugate to the $k'_{1 \mu \nu}$ and $k'_{2 \mu \nu}$ (massive graviton modes). 

To check this proposal holographicaly there are several suggestions. For example in \cite{Bergshoeff:2012ev} and for three dimensional tri-critical gravity the energy momentum tensor has been expanded in terms of the leading and sub-leading terms in Fefferman-Graham expansion of the metric and central charges have been computed. One may perform the same calculation for ${\tau_{1}}^{i}_{j}$ and ${\tau_{2}}^{i}_{j}$. We postpone the holographic study of RCG model for future works \cite{wip}.

\section{Schwarzschild-AdS black hole in RCG}
In this section we study $SAdS$ black hole as a solution of  the  RCG. First we study the thermodynamics of this black hole and then we compute its mass by renormalized boundary stress tensor. 

Let us start with the following $AdS_{d}$ black hole which is a solution of equations of motion (for simplicity we have considered a flat boundary space but it is possible to consider  spherical or  hyperboloid spaces)
\bea\label{SADS}
ds^{2}=\frac{l^{2}}{f(r) r^{2}} dr^{2}+ \frac{r^{2}}{l^{2}} \Big(-f(r) dt^{2}+ \sum_{a=1}^{d-2} \d_{ab}{dx^a}{dx^b}\Big)\,,\qquad f(r)=1-(\frac{r_0}{r})^{d-1}\,.
\eea
Here $r_0$ is the radius of the horizon and the cosmological parameter $\L_0$ is related to value of $l$, the radius of $AdS$ space-time, through the following relation
\be
\L_0\!=\! \frac{d-1}{2 l^6} \Big((d-1) (-(d-6) (d-1) (c_1+d (c_2+c_3 d))+(d-4) (a_1+a_2 d) l^2)-(d-2) l^4 \sigma\Big).
\ee
\subsection{Black hole thermodynamics}
To study the thermodynamics of this black hole we start from temperature and then compute the entropy. The value of temperature can be read from the Euclidean version of the metric by using the following relation
\be\label{TAdS}
T=\frac{1}{2\pi} \sqrt{g^{rr}} \partial_r(\sqrt{g_{\t\t}})\Big|_{r=r_0}=\frac{(d-1) r_0}{4\pi l^2}\,.
\ee
To find the entropy we employ two techniques, the free energy and the Wald's formula.
To compute the free energy we must insert the metric into the Euclidean action 
\begin{align}\label{hatsbar}
I^{BH}_E[T]&=-\frac{1}{16\pi G_d}\int_{0}^{1/T}\!\!\!d\tau \int_{r_0}^{R}\!\!\! dr\int d^{d-2}x \sqrt{g_E} \mathcal{L}
=-\frac{V_{d-2}}{2(d-1)G_d}\frac{r_0^{d-2}}{l^{d-2}}\Big(1-\frac{R^{d-1}}{r^{d-1}_0}\Big)\bar{\s}\,, 
\end{align}
where $\bar{\s}$ is given in equation (\ref{sigmas}) and we have considered $V_{d-2}$ as  the regulator volume of $d-2$ dimensional flat space. In above equation $R$ is a regulator for radial coordinate which we will send it to infinity later. To remove the divergent part of the above expression we need to subtract the value of action for $AdS$ background at temperature $T'$
\begin{align}
I^{AdS}_E[T']&=-\frac{1}{16\pi G_d}\int_{0}^{1/T'}\!\!\!d \tau \int_{0}^{R}\!\!\! dr\int d^{d-2}x \sqrt{g_E} \mathcal{L}=-\frac{V_{d-2}}{2(d-1)G_d}\frac{r^{d-1}}{r_0l^{d-2}}\Big(1-\frac{r_0^{d-1}}{R^{d-1}}\Big)^{\frac12}\bar{\s}\,, \nn \\ 
\end{align}
where $T'$ is defined in such a way that the time periodicity of the Euclidean $AdS$ background will be equal to the black hole's one at the regulator surface $r=R$. In other word
\be
\frac{1}{T'}=\frac{1}{T}\sqrt{\frac{g^{BH}_{tt}}{g^{AdS}_{tt}}}\Bigg|_{r=R
}=\frac{4\pi l^2}{r_0(d-1)}\,\Big(1-\frac{r_0^{d-1}}{R^{d-1}}\Big)^{\frac12}\,.
\ee
Finally the free energy is given by
\be\label{freeE}
F[T]=T(I^{BH}_E[T]-I^{AdS}_E[T'])\Big|_{R\rightarrow\infty}=-\frac{V_{d-2} r_0^{d-1}}{8\pi G_d l^{d}} \bar{\s}\,,
\ee
and the entropy can be read as
\be\label{entads}
S=-\frac{\partial F}{\partial T}= \frac{1}{2 G_d} (\frac{4\pi l}{d-1})^{d-2} T^{d-2} \bar{\s} \,.
\ee
It is worth to mention that the same result for entropy can be found from the well known Wald's formula for entropy in higher curvature theories of gravity \cite{Jacobson:1993vj}. Starting from
\begin{align}\label{Wald}
S^{W}&=8\pi \int_H d^{d-2} x^H \sqrt{g^H} g^{\perp}_{\a\m} g^{\perp}_{\b\n} \Big(\frac{\partial {\mathcal{L}}}{\partial R_{\a\b\m\n}}-
\nabla_{\g} \frac{\partial {\mathcal{L}}}{\partial\nabla_{\g} R_{\a\b\m\n}}\Big)\,,\nn \\
&=\frac{1}{2G_d} \int_H d^{d-2} x^H \sqrt{g^H} 
\Big(\tfrac{1}{2} \Bigl(g^{\perp}_{\alpha}{}^{\delta} g^{\perp}{}^{\alpha \beta} \bigl(-3 c_1 R_{\beta}{}^{\n} R_{\delta\n} - 2 R_{\beta\delta} (a_1 + c_2 R)\bigr) -  g^{\perp}_{\alpha \beta} g^{\perp}{}^{\alpha \beta} \bigl(\sigma \nn \\
&+ c_2 R^{\m\n} R_{\m\n} + R(2 a_2 + 3 c_3 R)\bigr) + g^{\perp}{}^{\alpha}{}_{\alpha} \Bigl(g^{\perp}{}^{\beta \delta} \bigl(3 c_1 R_{\beta}{}^{\n} R_{\delta\n} + 2 R_{\beta\delta} (a_1 + c_2 R)\bigr) + g^{\perp}{}^{\beta}{}_{\beta} \bigl(\sigma \nn \\
&+ c_2 R^{\m\n} R_{\m\n} + R(2 a_2 + 3 c_3 R)\bigr)\Bigr)\Bigr)-2 b_2 g^{\perp}{}^{\alpha}{}_{\alpha} g^{\perp}{}^{\beta \delta} \nabla_{\nu}\nabla^{\nu}R_{\beta \delta} - 2 b_1 g^{\perp}{}^{\alpha}{}_{\alpha} g^{\perp}{}^{\beta}{}_{\beta} \nabla_{\nu}\nabla^{\nu}R\Big)\,,
\end{align}
where $g^{\perp}_{\a\b}$ denotes the metric projection onto the subspace
orthogonal to the horizon, one can find the same entropy as (\ref{entads}) exactly.

\subsection{Mass from renormalized boundary stress tensor}
 The auxiliary field components can be determined by their field equations (\ref{lfeom1}) and (\ref{lfeom2}) in terms of the $SAdS$ black hole metric as follows
\bea\label{boundaux}
&&\hat{\lambda}^{i j}=-\frac{1}{2 l^{2}} \gamma^{i j}\,,\qquad \hat{\lambda}^{i r}= \hat{f}^{i r}=0 \,, \qquad \hat{\lambda}^{r r}=-\frac{1}{2l^2} \frac{r^{2}}{ l^{2}} f(r) \,.\nn\\
&&\hat{f}^{i j}=-\frac{2 (d-1) }{(d-2) l^{4}} \gamma^{i j} \Big(3 (d-1) (c_1 +d (c_2 + c_3 d)) - 2 (a_1 + a_2 d) l^2\Big)\,,\nn\\
&&\hat{f}^{r r}= -\frac{2 (d-1) }{(d-2) l^{4}} \Big(3 (d-1) (c_1 +d (c_2 + c_3 d)) - 2 (a_1 + a_2 d) l^2\Big) \frac{r^{2}}{ l^{2}} f(r)\,.
\eea
Using these relations the value of mass can be computed from equation (\ref{NRM}). To find the mass we need $T^{00}$ from (\ref{BST}) together with equations (\ref{ders}) and (\ref{hats}). The value of mass with this technique becomes
\begin{align}\label{infmass}
&M=\frac{(d-2)  V_{d-2}}{4\pi l G_d}  (\frac{r_0}{l})^{d-1}\Big(1-(\frac{r}{r_0})^{d-1}\Big) \bar\s \,,
\end{align}
which  diverges  obviously  when computed on boundary at $r\rightarrow\infty$. To remove this divergence we need to renormalize the energy-momentum tensor as follow
\be\label{tensor}
T^{ren}_{ij}=T_{ij}-\frac{d-2}{8\pi G_d l}  \bar\s \g_{ij}\,.
\ee
This can be done  by adding a boundary counter-term to the Lagrangian just proportional to the volume of boundary. Consequently the mass is given by
\begin{align}\label{sadsmass}
&M=\frac{(d-2) V_{d-2}}{8\pi l G_d}  (\frac{r_0}{l})^{d-1} \bar\s\,.
\end{align}
As a check of our results, it is easy to show that the value of  mass in (\ref{sadsmass}) and the values of entropy (\ref{entads}) and temperature (\ref{TAdS}) will satisfy the first law of thermodynamics for black holes i.e. $dM=TdS$, if one differentiates mass and entropy with respect to the location of the horizon at $r=r_0$. As another check one may compute the mass of $AdS$ black hole from the first approach by linearizing the equations of motion i.e. from equation (\ref{Qchargeb}). It can be shown that for asymptoticly $AdS$ solutions only the first term in  (\ref{Qchargeb}) has contribution to the AD mass.  This computation reconfirms the value of mass in equation (\ref{sadsmass}).
\section{Lifshitz vacuum and Lifshitz black hole in RCG}

In addition to the $SAdS$ black hole where we discussed in previous section, one can find other interesting solutions such as the Lifshitz vacuum and Lifshitz black hole \cite{Kachru:2008yh, Mann:2009yx , Bertoldi:2009vn,AyonBeato:2009nh, Dehghani:2010kd, Dehghani:2010gn,Brenna:2011gp,Giacomini:2012hg,Anastasiou:2013mwa,Brenna:2015pqa}.
It should be noted that the Lifshitz solutions  with different scaling of space and time  have interesting applications as gravity duals of non-relativistic quantum field theories.
In this section we will find conditions to have such solutions in RCG. We also use the auxiliary field formalism to compute the mass of Lifshitz black holes.

Let us start from the Lifshitz vacuum as a solution for equations of motion. This background is characterized by a dynamical exponent $z$, which governs the anisotropy between spatial and temporal scalings i.e. $x \rightarrow \lambda x, r \rightarrow \lambda^{-1} r$ and $t\rightarrow \l^z t$,
\bea
ds^{2}=-\frac{r^{2z}}{l^{2z}} dt^{2}+\frac{l^{2}}{ r^{2}} dr^{2}+ \frac{r^{2}}{l^{2}} (\sum_{a=1}^{d-2} \d_{ab}{dx^a}{dx^b}) \,.
\eea
Moreover the cosmological parameter $\L_0$ and the Lifshitz length $l$ in the metric above can be fixed by equations of motion and are given by 
\begin{dgroup}	
	\begin{dmath}\label{L0LS}
		\L_0=\frac{1}{l^6}\Big(-\frac12 (d-2) (d-1) l^4 \s+\tfrac{1}{2} l^2 ( z^2+d-2) \big(2 + d^2 - 2 (z-2) z -  d (3 
		+ 2 z)\big) a_1  +\tfrac{1}{2} l^2 \big((2 \\- 3 d + d^2)^2 - 4 (d-2)^2 z^2 - 8 (d-2) z^3 - 4 z^4\big) a_2
		+(2 -  d) (z-1)^2 z \big(12   - 5 z  + d (3z-14 \\+ 4 d)\big)b_2
		+\tfrac{1}{2} \big((d-2)^3(6z- (d-1)) - 6 (d-3) (d-2)^2 z^2   
		+ 2 (d-2) \big(5  + (d-4) d\big) z^3 \\+ 3 (d-2) (d-1) z^4 + 6 (d-2) z^5 + 4 z^6\big)c_1- \tfrac{1}{2} \big(d^2  + 2 (z-1)^2 + d (2z- 3)\big)^2 \big(2 + d^2 - 4 (z \\ \hiderel{}\,\,\,\, -2) z -  d (3 + 4 z)\big)\big(c_3+(z^2+d-2)c_2\big)\Big)\,, 
	\end{dmath}
	\
	\begin{dmath}\label{l2LS}
		l^2=\frac{1}{\s}\Big(a_1 (z^2+d-2) + a_2 \big(d^2 + 2 (z-1)^2 + d (2z-3)\big)\pm\frac12\sqrt{\d}\Big)\,, \\  
		\hiderel{} \!\!\!\!\!\!\!	\d  \hiderel{=} 4 \big(a_1 ( z^2+d-2) + a_2 \big(d^2 + 2 (z-1)^2 + d (2z-3)\big)\big)^2 - 4 \big((12  - 36 d + 39  d^2  
		- 18  d^3 + 3  d^4 \\+(- 48 + 96  d - 60  d^2  + 12  d^3) z + (72 - 84  d  + 24 d^2)z^2 +(- 48 	+ 24  d) z^3 + 12  z^4)c_3 + 3  \big(d^3 \\+ 2 (z-1)^2 (z^2-2 ) + d^2 (z^2+2z-5) + d (8 - 8 z -  z^2\big) 
		+ 2 z^3)c_2+ 3  ( z^2 +d -2)^2c_1  +( 2  d^2 z \\ \hiderel{} \,\,\,\, - 4  d z   + 6  z^2- 2 d^2 z^2 - 6  z^3  + 4  d z^3)b_2 \big) \sigma\,.
	\end{dmath}
\end{dgroup}
\subsection{Lifshitz black hole}
Motivated by $SAdS$ solution at $z=1$ one may choose the following ansatz to write the Lifshitz black hole 
\bea\label{LifBH}
ds^{2}=-\frac{r^{2z}}{l^{2z}} f(r) dt^{2}+\frac{l^{2}}{ r^{2} f(r)} dr^{2}+ \frac{r^{2}}{l^{2}} \sum_{a=1}^{d-2} \d_{ab}{dx^a}{dx^b}\,,\qquad f(r)=1-(\frac{r_0}{r})^{d-1}\,.
\eea
Similar to the Lifshitz background, the above metric is a solution for equations of motion in a special region on parameter space of the theory. These special values are  presented in equations (\ref{a1BHLS})-(\ref{L0BHLS}) in the appendix B. 

The auxiliary fields in this background are given by
\bea\label{auxlifbh}
&&{\l}^{ab}=- \frac{ \bigl((d-1)(d-2) + 2 z (1-z) + (3 - 3 d + 2 z) (z-1) (\frac{r_0}{r})^{d-1}}{2 (d-1)(d-2) }\frac{\d^{ab}}{r^2}\,,\nn \\ 
&&{\l}^{tt}=\frac{ \big(1 + 2 (z-2 ) z + d (2z-1)\big) - (z-1) (1 -d + 2 z) (\frac{r_0}{r})^{d-1}}{2 (d-1) l^2  \big(1- (\frac{r_0}{r})^{d-1}\big)}\frac{l^{2z}}{r^{2z}}\,, \nn \\  &&N^2 {\l}^{rr}=\frac{ - \big(d-1  + 2 (z-1) z\big) + (3-3 d + 2 z) (z-1) (\frac{r_0}{r})^{d-1}}{2 (d-1) l^2 } \,, \nn \\
&&{\l}^{ar}={\l}^{at}={f}^{ar}={f}^{at}=0\,,\\
&& {f}^{ab} = \frac{1}{2(d-2) l^2}\frac{\d^{ab}}{r^2}(f_{10}+f_{11}\frac{r_0^{d-1}}{r^{d-1}}+f_{12}\frac{r_0^{2d-2}}{r^{2d-2}})\,,\nn \\
&&
{f}^{tt}=\frac{1}{{4(d-2) l^4  (1-(\frac{r_0}{r})^{d-1})}}\frac{l^{2z}}{r^{2z}}(f_{20}+f_{21}\frac{r_0^{d-1}}{r^{d-1}}+f_{22}\frac{r_0^{2d-2}}{r^{2d-2}})\,, \nn \\
&&
N^2 {f}^{rr}=-\frac{1}{4(d-2)l^4 }(f_{30}+f_{31}\frac{r_0^{d-1}}{r^{d-1}}+f_{32}\frac{r_0^{2d-2}}{r^{2d-2}})\frac{l^2}{r^2 f(r)}\,,\nn
\eea
where all the constant values $f_{10}$ to $f_{32}$ are given in equations (\ref{f10})-(\ref{f32}).
\subsection{Thermodynamics of Lifshitz black hole}
With the same approach as $SAdS$ black holes we can read the value of temperature from Euclidean metric
\be\label{lifT}
T=\frac{d-1}{4\pi l} \frac{r_0^z}{l^z}\,.
\ee
To compute the entropy we suppose  that the Wald's entropy formalism still holds here and its value can be found by putting the black hole solution (\ref{LifBH}) into the Wald's formula which we have computed in equation (\ref{Wald})
\begin{align}\label{lifentropy}
S&=\frac{V_{d-2}}{8l^4G_d}\frac{r_0^{d-2}}{l^{d-2}}\tilde{\s}\,,\nn \\
\tilde{\s}&=(d-1) \Big((d-1)(3 c_1 (1-3z)^2+6 c_2(1 + 2 (d-5) z + 9 z^2) + 12 c_3  (d-3 + 3 z)^2\nn \\
& +16 (2b_1  (1 - d + 2 z) +  b_2  (3 - 2 d + 2 z)) (z-1))\Big)\nn \\
&+\frac{1}{l^2}\Big( a_1 (1 - 3 z)-2a_2 (d-3+3z)\Big) + 4l^4 \sigma\,.
\end{align}
Using the first law of thermodynamics for black holes, $dM=TdS$, we find that the mass is given by
\be\label{lifmass}
M=\frac{V_{d-2} (d-1)(d-2)}{32\pi l^5 G_d (z+d-2)}\frac{r_0^{z+d-2}}{l^{z+d-2}}\tilde{\s}\,.
\ee
As we see the value of the mass leads to the $SAdS$ black hole's mass (\ref{sadsmass}), when we insert $z=1$.

As a  general result, we observe that the  entropy (\ref{lifentropy}) and mass (\ref{lifmass}) for $SAdS$ or Lifshitz black holes in RCG, are both proportional to the critical parameter $\tilde \s$. Stability of these solutions restricts this critical parameter to a positive value i.e. $\tilde \s>0$. 

Several investigations in thermodynamical properties of the Lifshitz black holes have been done in other gravitational  theories, for example see \cite{ Cai:2009ac, AyonBeato:2010tm, Ayon-Beato:2015jga}. 
\subsection{Mass from the renormalized boundary stress tensor}
To find the mass of Lifshitz black hole from auxiliary field formalism we need to compute the integral in equation (\ref{NRM}) and the value of $T^{00}$ can be found by inserting the values of auxiliary field components in equation (\ref{auxlifbh}) into the relation  (\ref{BST}).
The non-renormalized black hole's mass in this way is given by
\begin{align}\label{divmass}
M &= - \frac{ V_{d-2}}{32\pi G_d\, l^{3 + d + z}  } (1 -  \frac{r_0^{d-1}}{r^{d-1}}) r^{ d+z-2}(  M_1 +  \frac{r_0^{ d-1}}{r^{ d-1}} M_2 + \frac{r_0^{2 d-2}}{r^{2 d-2}} M_3)  \\
& = - \frac{ V_{d-2}}{32\pi G_d\, l^{3 + d + z}  }  r^{ d+z-2} (  M_1 +  \frac{r_0^{ d-1}}{r^{ d-1}} (M_2-M_1) + \frac{r_0^{2 d-2}}{r^{2 d-2}} (M_3 -M_2)-M_3\frac{r_0^{3d-3}}{r^{3d-3}} )\,,\nn
\end{align}
where
\begin{align}
M_1  &=  -2 (d-2 ) f_{10} + f_{20} -  f_{30} + 8 (d-2) \bigl(2 b_2 (d-2) (z-1)^2 z + l^4 \sigma\bigr)\,, \nn \\
M_2 &= 2 f_{11} + f_{21} -  f_{31} +4 b_2  (z-1)^2(\big(4 + d (2d-5)\big) (d-1 - 2 z)-4(d-2)^2z) \nn \\
&+ 16 b_1 (d-1)^2 (d-1- 2 z) (z-1)^2 \,,  \\
M_3 &= 2 d f_{12} + f_{22} -  f_{32} - 4 b_2  (z-1)^2\big(4 + d (2d-5)\big) (d-1 - 2 z) \nn \\
& - 16 b_1 (d-1)^2 (d-1 - 2 z) (z-1)^2\,,\nn
\end{align}
and this mass diverges as one goes to the boundaries at $r\rightarrow\infty$. To have a finite non zero mass for Lifshitz black holes, the above relation suggests that the only possible massive black holes are those with $z=1, z=d$ and $z=2d-1$. The case with $z=1$ or $SAdS$ black hole has been studied already in the previous section . We now try to find the renormalized mass for two other cases.

Similar to $SAdS$ black hole  we need to find a renormalized boundary energy momentum tensor here. As noted in \cite{Hohm:2010jc}, there is an ambiguity in choosing the boundary terms. 
For example we can choose the following scalar tensors to construct the counter-terms on the boundary
\begin{align} 
&S_{ct}^{(1)}\!=\!\frac{1}{\k^2}\int d^{d-1} x \sqrt{-\g} \rightarrow
 M_{ct}^{(1)} = \frac{ V_{d-2}}{4\pi G_d\, l^{d + z-2}} (1 -  \frac{r_0^{d-1}}{r^{d-1}}) r^{ d+z-2}( 1 + \frac12 \frac{r_0^{ d-1}}{r^{ d-1}}  +\frac38 \frac{r_0^{2 d-2}}{r^{2 d-2}} ) \,, \nn \\
&S_{ct}^{(2)} \!=\! \frac{1}{\k^2}\int d^{d-1} x \sqrt{-\g}  \l^k{}_k \rightarrow
	M_{ct}^{(2)} =  -\frac{M_{ct}^{(1)}}{2l^2}(d+2z-3-2(z-1) \frac{r_0^{ d-1}}{r^{ d-1}}  )\,,
\end{align}
and etc. 
One may find different scalars in order to construct the renormalized action. For example if we restrict ourselves to at most cubic terms with at most two covariant derivatives we can choose the following scalars,
\begin{align}
&\Big\{f^k{}_k, f^r{}_r, \l^k{}_k, \l^r{}_r, f^i{}_i \l^j{}_j, f^k{}_k  \l^r{}_r, f^r{}_r \l^j{}_j, f^r{}_r  \l^r{}_r,  f^i{}_j \l^j{}_i,  \l^i{}_j \l^j{}_i  , (\l^k{}_k)^2,(\l^r{}_r)^2\nn \\
&, (\l^k{}_k)^3, (\l^r{}_r)^3, \l^i{}_i \l^j{}_k\l^k{}_j, \l^i{}_j \l^j{}_k\l^k{}_i, \l^i{}_i \l^r{}_r, 
 (\l^i{}_i)^2 \l^r{}_r,\l^i{}_i (\l^r{}_r)^2,\l^i{}_j \l^j{}_i \l^r{}_r \Big\}\,.\nn
\end{align}
Therefore to find a renormalized mass one encounters the ambiguity in choosing the correct combination of terms as indicated in \cite{Hohm:2010jc}. To find a renormalized mass we will follow the same steps as \cite{Hohm:2010jc} and fix the coefficients by using the value of mass in (\ref{lifmass}) which is consistent with the first law of thermodynamics for black holes.

 As an example let us start with the following combination of boundary  counter-terms which have been chosen for simplicity
\begin{align}\label{SCC}
S_{cc}=\frac{1}{\k^2}\int d^{d-1} x \sqrt{-\g}&(\a_1 +\a_2 \l^k{}_k +\a_3 (\l^k{}_k)^2 +\a_4 \l_{ij} \l^{ij} \nn \\
&+\a_5 (\l^k{}_k)^3+\a_6 \l_{ij} \l^{ij} \l^k{}_k+ \a_7 \l_{ij}\l^{jk}\l_{k}{}^i)\,.
\end{align}
By adding these counter-terms and demanding a finite value for mass equal to the value in  (\ref{lifmass}) for $z=d$ and $z=2d-1$ simultaneously, we can fix the unknown coefficients $\a_1$ to $\a_7$ which are given in the appendix C.

\section{Ricci cubic gravity in three dimensions}
In this section, as an application of our results, we are trying to study the RCG in three dimensions. We will compute the values of central charges corresponding to the dual CFT  of the $AdS_3$ space-time. We also find the BTZ black hole mass and angular momentum from the renormalized energy-momentum tensor.
\subsection{Central charges}
The central charges of the dual CFT of  $AdS_3$ space-time can be computed by applying the renormalized boundary stress tensor (\ref{BST}). We will review and use the method in  \cite{Balasubramanian:1999re} and \cite{Hohm:2010jc}.

To find the central charge we need the anomalous behavior of the energy-momentum tensor under the conformal transformation. In light-cone coordinates these transformations are 
\be \label{deltx}
\delta{x^{+}} =-\xi^{+}(x^{+})\,,     \qquad          \delta{x^{-}} =-\xi^{-}(x^{-})\,,
\ee
and consequently the energy-momentum tensor components transform as \cite{Balasubramanian:1999re} 
  \be\label{delt}
\delta{T_{++}}=\mathcal{L}_{\xi}T_{++}-\frac{c}{24\pi}\partial^{3}_{+}\xi^{+} , \qquad  \delta{T_{--}}=\mathcal{L}_{\xi}T_{--}-\frac{c}{24\pi}\partial^{3}_{-}\xi^{-}\,.
 \ee
Each transformation contains two parts, a Lie derivative part which comes from the boundary-preserving diffeomorphisms and an anomalous part which comes from the fact that the asymptotic symmetry group of $AdS_{3}$ is larger than the boundary-preserving diffeomorphisms.

To compute the central charges from the anomalous terms, let us start with the $AdS_{3}$ metric  written in the light-cone coordinates
\be
	ds^2=\frac{l^2}{r^2} dr^2-r^2 dx^{+} dx^{-}\,,
\ee
together with the Brown and Henneaux boundary conditions \cite{Brown:1986nw} to define the asymptotic behavior  of the  metric 
\begin{align}
	g_{+-}&=-\frac{r^2}{2} + \mathcal{O}(1)\,,\qquad g_{++}= \mathcal{O}(1)\,,\qquad g_{--}= \mathcal{O}(1)\,,\qquad \nn \\ 
	g_{rr}&=\frac{l^2}{r^2} + \mathcal{O}(\frac{1}{r^4})\,,\qquad g_{+r}= \mathcal{O}(\frac{1}{r^3}) \,,\qquad g_{-r}= \mathcal{O}(\frac{1}{r^3})\,.
\end{align}
The diffeomorphisms which respect to Brown and Henneaux boundary conditions are  parametrized by the following vector fields
\footnote{We note that in the ADM decomposition, the vectorial diffeomorphism parameter $X^{\mu}$ can be decomposed as $X^{\mu}=(\xi^{i} , \lambda)$
	where $\xi^{i}$ and $\lambda$  are arbitrary functions of coordinate  $x^{\mu}=(x^{i},r)$ \cite{Hohm:2010jc}.}
\be\label{asym}
	X^{\pm}=\xi^{\pm}(x^{\pm})+\frac{l^2}{2 r^2} \partial^{2}_{\mp} \xi^{\mp}(x^{\mp}) \,, \qquad 
	X^{r}=-\frac{r}{2} (\partial_{+}{\xi^{+}} + \partial_{-}{\xi^{-}})\,.
\ee
These diffeomorphisms (assymptotic symmetry group of $AdS_3$) do not belong to the class of boundary-preserving diffeomorphism \cite{Hohm:2010jc}  and therefore they will produce anomalous terms similar to those in (\ref{delt}).

To compute the transformation of boundarty energy momentum tensor we need to compute the transformation of the boundary metric, the extrinsic curvature and the auxiliary fields components under (\ref{asym}). For example we have 
\begin{align}
\delta_{X}{g_{\mu \nu}}=\mathcal{L}_{X}g_{\mu \nu}=X^{\rho}\partial_{\rho}g_{\m \nu}+ \partial_{\mu}X^{\rho}g_{\rho \nu}+ \partial_{\nu}X^{\rho}g_{\mu \rho} \,, \nn\\
\delta_{X}{\lambda^{\mu \nu}}=\mathcal{L}_{X}\lambda^{\mu \nu}=X^{\rho}\partial_{\rho}\lambda^{\m \nu}- \lambda^{\mu \rho} \partial_{\rho}X^{\nu}- \lambda^{\rho\nu} \partial_{\rho}X^{\mu}  \,,
\end{align}
Using these relations, all components of the metric remain invariant except two components $g_{++}$ and $g_{--}$ which transform as
\begin{align}\label{asyg}
\delta_{X}{g_{++}}=-\frac{l^2}{2} \partial^{3}_{+} \xi^{+} \,, \qquad \delta_{X}{g_{--}}=-\frac{l^2}{2} \partial^{3}_{-} \xi^{-}\,.
\end{align}
The extrinsic curvature $K_{ij}$ and its trace are also invariant and the only possible non-trivial components for  auxiliary fields (\ref{boundaux}) are (for more details of this computation see appendix A in  \cite{Hohm:2010jc})
\begin{align}\label{asyma}
	\delta_{X}{\lambda^{++}}&=-\frac{1}{ r^4} \partial^{3}_{-} \xi^{-}\,, \qquad \delta_{X}{\lambda^{--}}=-\frac{1}{r^4} \partial^{3}_{+} \xi^{+}\,, \nn\\
	\delta_{X}{f^{++}}&=-\frac{2}{ r^4} \big (a_{1} -\frac{30}{l^2}c_{1} -\frac{78}{l^2} c_{2} - \frac{216}{l^2} c_{3}+\frac{3}{l^2} b_{2}  \big)  \partial^{3}_{-} \xi^{-}\,,\nn\\
	\delta_{X}{f^{--}}&= -\frac{2}{ r^4} \big (a_{1} -\frac{30}{l^2}c_{1} -\frac{78}{l^2} c_{2} - \frac{216}{l^2} c_{3}+\frac{3}{l^2} b_{2}  \big)   \partial^{3}_{+} \xi^{+}\,.
\end{align}
where we have used the scalar ghost-free conditions in (\ref{a2b1}) to write the above relations. 

 Now we can use (\ref{asyg}) and (\ref{asyma}) to compute $\delta_{X}{T_{++}}$ from the renormalized energy-momentum tensor which we found in (\ref{tensor})
\be \label{T++}
8\pi G\delta_{X}{T_{++}}= -\frac{l}{2} \partial^{3}_{+} \xi^{+}(\sigma + \frac{1}{2 l^{2}} a_1-\frac{15}{l^{4} } c_1 -\frac{39}{l^{4}} c_2-\frac{108}{l^{4}} c_3+\frac{3}{2 l^{4}}  b_2)= -\frac{l}{2} \partial^{3}_{+} \xi^{+} \bar\sigma_{d=3}\,,
\ee
By comparing the above result with the second term in equation (\ref{delt}), one can read the central charge as
\be \label{Central}
 c=\frac{3l}{2G} \bar \sigma_{d=3}\,.
\ee
This value coincides with the value computed by another method in \cite{Afshar:2014ffa}.

An example of the three dimensional RCG is the Extended New Massive Gravity (ENMG) which is a theory free of scalar ghosts. It has been shown in \cite{Afshar:2014ffa} that, it is possible to write the action of ENMG in terms of three dimensional Schouten tensor, $\l_{\m\n}=R_{\m\n}-\frac14 R g_{\m\n}$  and Cotton tensor $C_{\m\n}={\epsilon^{\rho\kappa}}_{\nu} \nabla_{\rho}(\l_{\m\kappa})$ as
\begin{align}\label{ENMG}
\mathcal{L}=\sigma R -2 \Lambda_{0} +\frac{1}{m^{2}} (R_{\mu \nu}R^{\mu\nu} -\frac{3}{8} R^{2} ) +\frac{1}{m^{4}}(2 a\, det(\l)- b\, C^{\mu \nu}C_{\mu \nu}) \,,
\end{align}
where
\begin{align}
& det(\l)= -\frac13 {R^{\nu}}_{\mu} {R^{\rho}}_{\nu} {R^{\mu}}_{\rho} + \frac{3}{8} R R_{\mu\nu} R^{\mu \nu} - \frac{17}{192} R^{3}\,, \nn\\
& C^{\mu \nu}C_{\mu \nu}=R_{\mu \nu} \Box R^{\mu \nu} - \frac{3}{8} R \Box R -3 {R^{\nu}}_{\mu} {R^{\rho}}_{\nu} {R^{\mu}}_{\rho} + \frac{5}{2} R R_{\mu\nu} R^{\mu \nu} - \frac{1}{2} R^{3} \,.
\end{align}
Simply one can  match the two actions in (\ref{action}) and in (\ref{ENMG}) as follow 
\begin{align}\label{couplings}
& a_{1}=\frac{1}{m^{2}} \,, \quad a_{2}=-\frac{3}{8 m^{2}}\,, \quad 
 b_{1}=-\frac{3 b}{8m^{4}} \,, \quad  b_{2}=\frac{b}{m^{4}} \,, \quad \nn\\
& c_{1}=\frac{9b-2 a}{3 m^{4}}\,, \quad 
 c_{2}=\frac{3 a-10b}{4 m^{4}}\,, \quad  c_{3}=\frac{48b-17 a}{96 m^{4}}\,.
\end{align} 
By substitution of these values into the definition of $\bar \sigma_{d=3}$ in (\ref{sigmas}), the central charge (\ref{Central}) will be equal to  the central charge reported in \cite{Afshar:2014ffa}, i.e.
\begin{align}
c_{ENMG}= \frac{3l}{2G} \bar \sigma_{d=3}=\frac{3l}{2G} (\sigma + \frac{1}{2 m^2 l^{2}}  -\frac{a}{8 m^4 l^{4} } )\,.
\end{align} 

As an alternative way, the values of central charges can be computed from  the conserved charges  of (\ref{Qchargeb}). This method has been used  in \cite{Bergshoeff:2012ev} and \cite{Ghodsi:2011ua}, which again confirms the above value for central charges
\be
c_{L}= c_{R}=\frac{3l}{2G} \bar \sigma\,.
\ee

\subsection{BTZ black hole}
The BTZ black hole is a solution of pure gravity \cite{Banados:1992wn}. Here for RCG we have such a solution again. Starting from 
\bea
ds^{2}={l^2} \Big[\frac{r^2}{(r^2-r_{+}^2) (r^2-r_{-}^2)} dr^{2} - \frac{(r^2-r_{+}^2) (r^2-r_{-}^2) }{r^2} dt^{2} + r^{2} (d\phi- \frac{r_{+} r_{-}}{r^2} dt)^{2} \Big]\,,
\eea
where $r_{\pm}$ are outer and inner event horizons, the ADM-like metric for  rotating BTZ black hole  is given by
\begin{align}\label{BTZm}
ds^{2}&=N^{2} dr^{2} - N^{-2} dt^{2} + r^{2} (d\phi- N_{\phi} dt)^{2}\nn \\
N&=(-8GM+ \frac{r^{2}}{l^{2}}+\frac{16G^{2} J^{2}}{r^{2}})^{-\frac{1}{2}}\,, \qquad  N_{\phi}=-\frac{4GJ}{r^{2}}\nn \\
M&=\frac{r_{+}^2+r_{-}^2}{8 G l^2}\,, \qquad J=\frac{r_{+} r_{-}}{4G l}\,.
\end{align}
We can use this metric to compute the conserved charges from (\ref{Qzeta})  by using the stress tensor in (\ref{BST}) and after renormalization. Then the mass and angular momentum become 
 \bea \label{BTZmassang}
 M_{BTZ}=M\bar{\s}_{d=3}\,,\qquad  J_{BTZ}=J \bar\s_{d=3}\,.
 \eea
Moreover the angular velocity at horizon is defined as
\bea\label{BTZV}
\Omega_{H}= \frac{1}{l}N_{\phi}(r_{+})=\frac{1}{l} \frac{r_{-}}{r_{+}}\,.
\eea
To find the thermodynamical parameters of the BTZ black hole we observe that the Hawking temperature in ADM form of the metric can be obtained  from the surface gravity $\kappa$ as 
\bea\label{BTZT}
T_{H}=\frac{\kappa}{2 \pi}=\frac{1}{2 \pi l} \frac{\partial_{r}{N}}{\sqrt{g_{r r}}}=\frac{r_{+}}{2 \pi l} (1- \frac{r_{-}^2}{r^2_{+}}) \,.
\eea
Now by using the Smarr relation $M=\frac{1}{2} T_{H} S_{BH}+\Omega_{H} J$ for BTZ black holes in three dimension we can evaluate the entropy. The Bekenstein-Hawking entropy is given by
\bea\label{BTZS}
S_{BH}=\frac{\pi r_{+} {\bar\s}_{d=3}}{2l G}\,.
\eea
As we see, all conserved charges such as mass, angular momentum and entropy are proportional to the central charge of the dual CFT, therefore the BTZ black hole is stable whenever this central charge is positive. 
\section{Summary and conclusion}
In this paper we have studied the most general Ricci Cubic Gravity (RCG) in $d$ dimensional space-time. Our Lagrangian in (\ref{action}) is constructed out of the Ricci tensor up to cubic terms and its covariant derivatives such that the equations of motion only contain at most six partial derivatives. We have also considered a cosmological parameter. Since we are interested to study this theory from the auxiliary field formalism point of view, we have restricted ourselves to Ricci tensors. As has been shown in  \cite{Hohm:2010jc}, in three dimensions one needs to consider  two rank-two auxiliary fields to construct a Lagrangian with at most two derivatives in its equations of motion. A similar situation holds in general $d$ dimensions. 

If we add terms including the Riemann tensor, then we will need to consider rank-four auxiliary field \cite{Hohm:2010jc}, where we have postponed study of these terms for future works.
Our study of RCG is divided into two main parts: 

$\bullet$ The linear excitations of gravitational fields around the maximally symmetric $AdS_d$ space-time: We have used two different approaches to study the gravitons in RCG. In first approach in section two,  we linearize the equations of motion, following to the work of \cite{Bergshoeff:2012ev}, to write these equations as (\ref{lineom}). To get ride of scalar ghosts in this theory it would be enough to set the trace of linearized equations of motion to zero. Doing this, we will find two constraints among the nine free parameters (couplings and cosmological parameter) in this theory, (\ref{a2b1}). 

The scalar ghost-free condition allows us to use  the transverse-traceless gauge so that, the linearized equation of motion now can be decomposed as equation (\ref{3Box}). This suggests the existence of three exciting modes in $AdS_d$ background, two massive gravitons in addition to a massless one. Although seven free parameters have been remained in the parameter space of the theory but the mass of  massive modes, $(M_\pm)$, depend only on three parameters $\{\bar\s, \s_1,\s_2\}$ in (\ref{mpm}). These parameters have already defined in (\ref{sigmas}) as a linear combination of parameters of the theory. The stability of the theory in this background (tachyon-free condition or $M^2_\pm\geq 0$) suggests that the allowed regions of parameter space are restricted. We have summarized our results in table \ref{tab:table1}. Similar to the three dimensional case, as discussed in \cite{Bergshoeff:2012ev}, here we have also special subspaces in our three-parameter space $\{\bar\s, \s_1,\s_2\}$ where we have two or three degenerate massless gravitons or two degenerate massive gravitons.

In second approach in section three, we employ the auxiliary field formalism which has been introduced in  \cite{Hohm:2010jc}. To find the graviton mass spectrum we consider the excitation modes around the background metric $g_{\m\n}$ and the auxiliary fields $(f_{\m\n}, \l_{\m\n})$ and we show that, similar to three dimensions \cite{Bergshoeff:2012ev},
we can find a  linear combination of three Pauli-Fierz  spin-two Lagrangians (\ref{L2D}). The mass spectrum in this way confirms the results of the first approach. 

On the other hand, the second approach shows that for $\bar{\sigma}\neq 0$, in general it is impossible to avoid the rank-two ghost fields. Our computations confirm the known observation for pure gravitational theories with higher curvature terms that, we can not have both tachyon-free condition and ghost-free condition simultaneously. This statement can be verified by computing the  energy of the linear excitations too. This obtains from the Hamiltonian formalism  and by comparing the overall signs of energies. The results are given in equations (\ref{Egraviton0}) and (\ref{Egraviton1}).

$\bullet$  Black hole solutions and conserved charges: The RCG as a theory of gravity with higher curvature terms admits different solutions for equations of motion. In this paper we have focused on two types of solutions: The Schwarzschild $AdS$ (\ref{SADS}) and Lifshitz black holes (\ref{LifBH}). In section five we have investigated different properties of $SAdS$ solution such as the mass, Hawking temperature and entropy. 

The mass has been computed in two different ways. In section (3.2) we first use the Abbot-Deser method  \cite{Deser:2002jk} to find the conserved charges corresponding to the symmetries of the solution. This can be done by using the linearized equation of motion following \cite{Bergshoeff:2012ev} and the conserved charge is given  by equation (\ref{Qchargeb}). We can use it to compute the mass of black hole simply by considering a time-like Killing vector. 

On the other hand we can also compute the conserved charges in auxiliary field formalism. This can be done by computing the boundary stress tensor in this formalism. The energy-momentum tensor can be found by variation of the action with respect to the auxiliary fields and metric. To have a well-defined variational principle we need a generalized Gibbons-Hawking term (\ref{GH}). The final result is presented in equation (\ref{BST}) and mass can be computed from (\ref{NRM}). 

The value of mass in this way diverges, as one goes to the boundary at $r\rightarrow\infty$, see equation (\ref{infmass}). To find a finite answer, we must renormalize the boundary terms by adding some proper counter-terms. The final value of mass in this way, is given in equation (\ref{sadsmass}) and agrees with the mass from the first approach. 

We have studied the thermodynamical properties of $SAdS$ black holes  in section five, where we have found the entropy of the black hole both by direct computation of the free energy (\ref{entads}) and the Wald's entropy formula (\ref{Wald}). The values of mass, temperature and entropy satisfy the first law of thermodynamics for black holes, i.e. $dM=T dS$.

To complete our analysis for more complicated cases, in section six we study the Lifshitz black hole and try to compute its mass from the boundary stress tensor which we found from the auxiliary field formalism. To have such a solution we need to restrict ourselves to the special values in parameter space of the RCG. In this case all constants can be written in terms of two constants, $b_2$ and $\sigma$ of the parameter space and also the  dynamical exponent $z$ (see appendix B). 

Computing the mass, again gives a divergent answer (\ref{divmass}) but unlike the $SAdS$ black hole it contains four different divergent behaviors when one goes to the boundary. We show that in order to have a finite massive Lifshitz black hole, we have just three options for dynamical exponent, $z=1$ or $SAdS$,  $z=d$ and $z=2d-1$. 

As it has been noted in \cite{Hohm:2010jc} for three dimensional Lifshitz black holes, there is an ambiguity in choosing the counter-terms to renormalize the boundary terms. In our study this happens again and there are various possibilities to have a finite mass. Although it is not known that the Wald's entropy formula works here for Lifshitz black holes but one can use it naively to find a finite mass from the validity of the first law of thermodynamics for Lifshitz black holes. The value of this mass is given in equation (\ref{lifmass}). We fix the coefficients of the counter-terms (\ref{SCC}) on the boundary  so that the value of the mass is equal to its value in equation (\ref{lifmass}) (see appendix C). We should note that we will recover the $SAdS$ results at $z=1$.

In section seven, as an application of our results, we have studied the three dimensional RCG. For example we have computed the central charge associated to the dual CFT of $AdS_3$ space-time. We have also calculated the mass and angular momentum of BTZ black holes. Our results confirm the known results in the literature when one considers special values of parameters in NMG or ENMG theories. 
By looking at the values of mass and  entropy of BTZ black hole one can show that the stability condition holds when $\bar\s_{d=3}>0$. This coincides with the unitarity condition of the dual CFT, because the value of central charge is also proportional to $\bar\s_{d=3}$.

As a  general result in $d$ dimensional space-time, we observe that the  entropy (\ref{lifentropy}) and mass (\ref{lifmass}) for those black holes in RCG where we have considered in this paper, are both proportional to a specific parameter $\tilde \s$ and stability of solutions requires that this critical parameter of the theory must have a positive value i.e. $\tilde \s>0$. For $z=1$ the value of $\tilde{\s}$ reduces to $\bar{\s}$ for $SAdS$ black holes. A similar behavior has been  already reported for Gauss-Bonnet gravity  in \cite{Cvetic:2001bk}.

There are some open questions which we have postponed  for  further works  \cite{wip}:

1. It would be interesting to solve the ambiguity in choosing the counter-terms which renormalize the boundary stress tensor. Our choice for these counter-terms in (\ref{SCC}) is motivated by the holographic renormalization (see for example \cite{deHaro:2000vlm}). As we mentioned before, we have fixed the coefficients in auxiliary field formalism so that $\lambda_{\m\n}$ becomes the Schouten tensor in $d$ dimensions. By replacing (\ref{lfeom1}) into the equation  (\ref{SCC}) we will have a Lagrangian with counter-terms constructed out of the Ricci tensors alone. These terms have been made out of the induced metric on the boundary. It would be interesting to built such a Lagrangian by the method of holographic renormalization and then translate it to the auxiliary field formalism.

2. One can consider the contribution of the Riemann tensor into our analysis. But as indicated in \cite{Hohm:2010jc} it needs to introduce a rank-four auxiliary field into the game. This will make the analysis more complicated due the existence of total derivative terms such as Gauss-Bonnet term \cite{wip}.

      

\appendix
\section{Lifshitz parameters}

Coefficients related to the auxiliary fields in equation (\ref{auxlifbh})
\begin{dgroup}
	\begin{dmath}\label{f10}
	\\ \hiderel{}	\!\!\!\!\!\!\\ \hiderel{}	\!\!\!\!\!\!
		f_{10}= 4 a_1 l^2 \bigl(d-2 + (d-2) z + 2 z^2\bigr) + 8 a_2 l^2 \bigl(d^2 + 2 (z-1)^2 + d (2z-3)\bigr) +8 b_2 (d-2) (z-1) z \\
		- 6 c_1 \bigl((d-2)^2 + (d-2) d z^2 + 2 (d-2) z^3 + 2 z^4\bigr) - 4 c_2 \bigl(d^2 + 2 (z-1)^2 + d (2z-3)\bigr) \bigl(-4 \\
		\hiderel{}\,\,\,\,\, + d (2 + z) + z (3z-2)\bigr) - 12 c_3 \bigl(d^2 + 2 (z-1)^2 + d (2z-3)\bigr)^2\,,
	\end{dmath}
\begin{dmath}\label{f11}
	\\ \hiderel{}	\!\!\!\!\!\!\\ \hiderel{}	\!\!\!\!\!\!
	f_{11}=(z-1) \big( 4 a_1 l^2 (2d-3 - 2 z)+8 a_2 l^2 (d-1 - 2 z)  - 8 b_1 (d-1) (d-1 - 2 z) (d -  z) + b_2 \big(4 \\- 2 d (1 + d) + 20 z + 8 (d-2) d z - 8 (d-1) z^2\big) \!-\! 6 c_1 \big(10 + d (3d-11) + 6 z  + (d -5) d z	
	+ 2 (d \\ -1) z^2 - 4 z^3\big)  - 4 c_2 \bigl(-2 + d \big(7 + d (2d-7)\big) + 24 z + 4 (d-5) d z
	+ 2 (1 + d) z^2  - 12 z^3\big)\\ 	
	\hiderel{}\,\,\,\,\,+ 24 c_3 \big((2 -  d) (d-1)^2 + 2 (d-3) z^2 + 4 z^3\big)\,,
\end{dmath}
\begin{dmath}\label{f12}
	\\ \hiderel{}	\!\!\!\!\!\!\\ \hiderel{}	\!\!\!\!\!\!
	f_{12}=(z-1) \big(8 b_1 (d-1) (d-1 - 2 z) (2d-1 -  z) + 2 b_2 \big(-8 + d \big(17 + d (4d-13)\big) - 6 z  + 4 (3 \\- 2 d) d z + 4 (d-1) z^2\big)  - 3 c_1 (z-1) \big(13 + 5 d^2 - 8 d (2 + z) + 4 z (3 + z)\big) - 2 c_2 (z-1) \big(9 d^2 \\ 	\hiderel{}\,\,\,\,\,- 4 d (6 + 5 z) + (3 + 2 z) (5 + 6 z)\big)- 12 c_3 (d-1- 2 z)^2 (z-1)\big)\,, \\
\end{dmath}
\begin{dmath}\label{f20}
	\\ \hiderel{}	\!\!\!\!\!\!\\ \hiderel{}	\!\!\!\!\!\!
		f_{20}=4 \big( 2 a_1 l^2 \big(-2 + d^2 (z-1) - 4 (z-2) z + d \big(3 + (z-6) z\big)\big)- 4 a_2 l^2 \big(d^2 + 2 (z-1)^2 + d (2z-3)\big)\\ +4 b_2 (d-2)^2 (z-1) z^2 + 3 c_1 \big((d-2)^2 (d-1) + 2 (d-2)^2 z -  (d-2) \big(3 + (d -5) d\big) z^2 \\- 2 (d-3) (d-2) z^3 -  (d-4) z^4\big) - 2 c_2 \big(d^2 + 2 (z-1)^2 + d (2z-3)\big) \big(d^2 (z-1)+ (8 - 5 z) z  \\ 	\hiderel{} \,\,\,\,\,+ d \big(2 + (z-6) z\big)\big) + 6 c_3 \big(d^2 + 2 (z-1)^2 + d (2z-3)\big)^2\big) \,,
	\end{dmath}
\begin{dmath}\label{f21}
	\\ \hiderel{}	\!\!\!\!\!\!\\ \hiderel{}	\!\!\!\!\!\!
	f_{21}=2 (z-1) \big( 2 a_1 (d-4) l^2 (d-1 - 2 z) -8 a_2 l^2 (d-1 - 2 z) + 8 b_1 (d-1) (d-1- 2 z) \big(1  + (d-2) z\big) \\+ 2 b_2 (d-1- 2 z) \big(5d-4 -  d^2 + (d-2) (5d-8) z\big) + 6 c_1 \big(2 - 3 d + d^2 -  (d-2) \big(7 + (d-4) d\big) z\\ + \bigl(4 + (d-3) d\big) z^2 + 2 (d-4) z^3\big)+ 2 c_2 \big(-8 -  d \big(-26 + d \big(27 + (d-10) d\big)\big) + 28 z - 2 d \big(15 \\+ (d-6) d\big) z + 4 \big(11 + (d-6) d\big) z^2 + 8 (d-5) z^3\big) + 24 c_3 (d-1 - 2 z) \big(d^2 + 2 (z-1)^2 + d (2z\\ \hiderel{} \,\,\,\,\,\,-3 )\big) \,,
\end{dmath}
\begin{dmath}\label{f22} 
	\\ \hiderel{}	\!\!\!\!\!\!\\ \hiderel{}	\!\!\!\!\!\!
	f_{22}=(z-1) \big(8 b_1 (d-1) (d-1 - 2 z) \big(1 + d^2 + 4 z - 2 d (2 + z)\big)+ 2 b_2 (d-1 - 2 z) \big(-4 (1 + 4 z) \\+ d \big(17 + d (4d-17 - 6 z) + 20 z\big)\big) - 3 c_1 (z-1) \big((d-11) (d-3) d - 4 (d-7) d z + 4 (d-4) z^2 \\- 4 (5 + 8 z)\big)  - 4 c_2 (z-1) \big(-13 + d^3 -  d^2 (11 + 4 z) - 4 z (7 + 5 z) + d \big( 4 z (7 + z) +23\big)\big) \\ 	\hiderel{} \,\,\,\,\, +24 c_3 (d-1 - 2 z)^2 (z-1) \big)\,,
\end{dmath}
	\begin{dmath}\label{f30}
		\\ \hiderel{}	\!\!\!\!\!\!\\ \hiderel{}	\!\!\!\!\!\!
		f_{30}=4 \big( 2 a_1 l^2 \big(2 - 4 (z-1) z + d \big( (z-2) z-1\big)\big) - 4 a_2 l^2 \big(d^2 + 2 (z-1)^2 + d (2z-3 )\big) -4 b_2 (d\\ -2)^2 (z-1)^2 z- 3 c_1 (d-2 + z^2) \big(2 - 4 (z-1) z + d \big(-1 + (z-2) z\big)\big) - 2 c_2 \big(d^2 + 2 (z-1)^2 + d (2z-3)\big) \big(4 + (4 - 5 z) z + d \big(-2 + (z-2) z\big)\big)  + 6 c_3 \big(d^2 + 2 (z-1)^2 + d (2z-3)\big)^2\big)\,, \\
	\end{dmath}
\begin{dmath}\label{f31}	
\\ \hiderel{}	\!\!\!\!\!\! \\ \hiderel{}	\!\!\!\!\!\!
	f_{31}=2 (z-1) \big( 2 a_1 l^2 \big(12 + d (3d-13 -	2 z) + 8 z\big)-8 a_2 l^2 (d-1 - 2 z) - 8 b_1 (d-1) (d  -1 -	2 z) \big((d\\-2) d + z\big)  - 2 b_2 \big(8 + 4 (d-3) d + 12 z + d \big(-1 +	(d-4) d\big) z - 2 \big(12  + d (3d  -11)\big) z^2\big)  + 6 c_1 \big(22 \\ + d \big(-35 - 3 (d-6)d\big) + 18 z + 3 (d-5) d z -  (d-3) (3d -4) z^2 + 2 (d-4) z^3\big)+ 2 c_2 \big(-16 \\+ d \big(50 + d \big(-53 + (22 - 3 d) d\big)\big) + 68z - 2 d \big(39 + d (2d  -15)\big) z - 4 (d -3) (1 + d) z^2 + 8 
	(d \\ 	\hiderel{} \,\,\,\,\, -5) z^3\big)  - 24 c_3 \big((2 -  d) (d-1)^2 + 2 (d  -3) z^2 + 4 z^3\big)\big)\,, \\
\end{dmath}
\begin{dmath}\label{f32}
\\ \hiderel{}	\!\!\!\!\!\! \\ \hiderel{}	\!\!\!\!\!\!
	f_{32}=(z-1) \big( 8 b_1 (d-1) (d-1 - 2 z) \big(3 + d (3d-8) + 2 z\big) + 2 b_2 \big(28 + (d-3) d \big(21 + d (4d -11)\big) \\+ 2 d (4 + d -  d^2) z - 4 \big(4 + (d-3) d\big) z^2\big) - 3 c_1 (z-1) \big(9 d^3 - 2 d^2 (29 + 6 z) - 4 \big(17 + 4 z (4 \\+ z)\big) + d \big(113 + 4 z (15 + z)\big)\big) - 4 c_2 (z-1) \big(3 (d-1) \big(7 + (d-6) d\big) - 4 \big(11  + d (2d-11 )\big) z  \\ 	\hiderel{} \,\,\,\,\,+ 4 (d-5) z^2\big) +24 c_3 (d-1 - 2 z)^2 (z-1)\big)\,,
\end{dmath}
\end{dgroup}

\pagebreak
\section{Lifshitz couplings}
The value of couplings are fixed by equations of motion for Lifshitz black hole

\begin{dgroup}
\begin{dmath}
	b_2=-\frac{2 l^2 (d-2)(z-1) (d- 1 - 2 z)^2}{ \Xi\, (d- 1)}  \, \Big(b_1 (d- 1 )^2 \Big(144 \, d^8 + d^7 (- 1349 + 245\, z) + d^6 (5069  \\ \hiderel{}\!\!\!\!\!\!\!\!\!- 1681 \,z - 28 \,z^2) + d^5 (- 9631 + 4214 \, z + 503 \, z^2 - 366\, z^3) - 2 d^4 (-\, 4699+ 2221 \,z + 1227 \, z^2  \\ \hiderel{}\!\!\!\!\!\!\!\!\! - 1098 z^3 + 69 z^4)  + d^3 (- 3841 + 1493 z + 5384 z^2 - 6100 z^3 + 704 z^4 + 296 z^5) + d^2 (- \!229 - 269 z \\ \hiderel{}\!\!\!\!\!\!\!\!\! 
	- 5798 z^2 + 9584 z^3 - 2028 z^4 - 1284 z^5 + 408 z^6) + d (421 + 1128 z + 2425 z^2 - 6150 z^3 + 536 z^4  \\ \hiderel{}\!\!\!\!\!\!\!\!\!  + 2408 z^5 - 640 z^6 - 128 z^7) - 2 (- 9  + 344 z - 176 z^2 + 422 z^3 
	- 1775 z^4 + 1398 z^5 + 172 z^6 - 504 z^7  \\ \hiderel{}\!\!\!\!\!\!\!\!\! + 128 z^8)\Big)+ 2 l^4 (z  - 1) \Big(- 36 d^5 + 39 d^4 (5 + 3 z) - 3 d^3 (134 + 169 z + 33 z^2) + d^2 (399 + 789 z 
	+ 316 z^2  \\ \hiderel{}\!\!\!\!\!\!\!\!\!  + 8 z^3) + d (-192   - 549 z - 319 z^2 - 20 z^3 + 12 z^4) + 2 (18 + 75 z + 69 z^2 - 22 z^3 - 4 z^4 + 8 z^5)\Big) \sigma\Big) \,,
\end{dmath}

\begin{dmath}\label{a1BHLS}
a_1 = \frac{2}{\Xi} (2z- d+ 1) \Big(b_1 (d- 1)^2 \Big(96 d^{12} - 4 d^{11} (313 + 155 z) + d^{10} (5748 + 11168 z - 2084 z^2) \\ \hiderel{}\!\!\!\!\!\!\!\!\! - 2 d^9 (2380 + 43717 z - 14584 z^2 + 623 z^3)- 2 d^8 (29538 - 194013 z + 83251 z^2 -  2559 z^3 - 1681 z^4)\\ \hiderel{}\!\!\!\!\!\!\!\!\!+ d^7 (283749 - 1064198 z + 479036 z^2 + 59252 z^3 - 57337 z^4 + 5306 z^5) + d^6 (- 639651  + 1830052 z \\ \hiderel{}\!\!\!\!\!\!\!\!\! - 618548 z^2 - 557360 z^3 + 376829 z^4 - 58148 z^5 + 858 z^6) - 2 d^5 (- 420065 + 932631 z + 98489 z^2 \\ \hiderel{}\!\!\!\!\!\!\!\!\!- 999191 z^3 
+ 623451 z^4 - 105002 z^5 - 8679 z^6 + 2422 z^7) - 4 (z- 1)^3 (729 + 2043 z - 18693 z^2 \\ \hiderel{}\!\!\!\!\!\!\!\!\!+ 43087 z^3 - 23934 z^4 - 2440 z^5 + 5912 z^6 - 1776 z^7 + 256 z^8) - 2 d^4 (323943 - 431617 z - 920827 z^2  \\ \hiderel{}\!\!\!\!\!\!\!\!\!\! + 1885347 z^3 - 1080780 z^4 + 88096 z^5 + 104600 z^6 - 27774 z^7 + 1196 z^8)  + d^3 (258053 
+ 198738 z \\ \hiderel{}\!\!\!\!\!\!\!\!\! - 2573016 z^2 + 3774132 z^3 - 1580157 z^4 - 706838 z^5 + 801404 z^6 - 216300 z^7 + 13088 z^8  + 368 z^9) \\ \hiderel{}\!\!\!\!\!\!\!\!\! + d^2 (- 22947 - 403568 z + 1508244 z^2 - 1496736 z^3 - 571491 z^4 + 2071828 z^5 - 1423398 z^6  \\ \hiderel{}\!\!\!\!\!\!\!\!\!+ 348044 z^7 + 2936 z^8 - 10224 z^9 + 768 z^{10}) + 2 d (- 7560 + 65220 z- 106625 z^2 - 206748 z^3 \\ \hiderel{}\!\!\!\!\!\!\!\!\!+ 862502 z^4  - 1090236 z^5 + 606427 z^6 - 96372 z^7 - 45296 z^8 + 21872 z^9 - 3440 z^{10} + 256 z^{11})\Big) \\ \hiderel{}\!\!\!\!\!\!\!\!\! + l^4 (z- 1) \Big(- 48 d^9 
+ 6 d^8 (73 + 91 z) -  d^7 (1067 + 6051 z + 1090 z^2) + d^6 (- 2216 + 26887 z \\ \hiderel{}\!\!\!\!\!\!\!\!\!+ 12187 z^2  + 126 z^3) + d^5 (17779 - 59698 z - 57313 z^2 - 670 z^3 
+ 206 z^4) + 2 d^4 (- 21109 + 31661 z \\ \hiderel{}\!\!\!\!\!\!\!\!\! + 71906 z^2 + 3861 z^3 - 3173 z^4 + 410 z^5) -  d^3 (- 51745 
+ 11323 z + 203100 z^2 + 38588 z^3 - 30914 z^4 \\ \hiderel{}\!\!\!\!\!\!\!\!\!
+ 3736 z^5 + 296 z^6) + d^2 (- 34574 - 40589 z + 155707 z^2 + 81754 z^3 - 59310 z^4 + 3716 z^5 + 2824 z^6 \\ \hiderel{}\!\!\!\!\!\!\!\!\!- 304 z^7)  - 2 (711 + 5151 z - 2031 z^2
- 15167 z^3 + 8904 z^4 + 1520 z^5 - 1984 z^6 + 240 z^7 + 64 z^8) \\ \hiderel{}\!\!\!\!\!\!\!\!\!+ d (11583 + 37208 z - 54777 z^2 - 79318 z^3 + 51256 z^4 + 2336 z^5 - 6368 z^6  + 864 z^7 + 64 z^8)\Big) \sigma\Big)\,,
\end{dmath}
\begin{dmath}
a_2=\frac{1}{\Xi} \Big(b_1 (d- 1)^2 \Big(96 d^{12} (1 + z) - 4 d^{11} (373 + 492 z + 215 z^2)+ 2 d^{10} (4439+ 10435 z + 5265 z^2  \\ \hiderel{}\!\!\!\!\!\!\!\!\! + 309\, z^3) + 2 d^9 (- \,9625 - 70978\, z - 24276\, z^2 - 4618\, z^3 + 729\, z^4) + d^8 (- 45851  + 648636\, z  \\ \hiderel{}\!\!\!\!\!\!\!\!\! + 88120 z^2 + 52314 z^3 - 18061 z^4 + 2330 z^5) - 2 d^7 (- 213422 + 1010682 z - 20359 z^2 + 62966 z^3  \\ \hiderel{}\!\!\!\!\!\!\!\!\! - 47134 z^4 + 13244 z^5 + 999 z^6) + d^6 (- 1351601 + 4315236 z - 407200 z^2 + 7950 z^3- 271511 z^4  \\ \hiderel{}\!\!\!\!\!\!\!\!\! + 102462 z^5 
+ 44600 z^6 - 6720 z^7) + d^5 (2505650 - 6248772 z + 411834 z^2 + 756888 z^3 + 398394 z^4  \\ \hiderel{}\!\!\!\!\!\!\!\!\! - 70176 z^5 - 321534 z^6 + 68196 z^7 - 896 z^8) + d^4 (- 2971401 + 5919268 z + 764380 z^2 - 2348514 z^3 \\ \hiderel{}\!\!\!\!\!\!\!\!\!  \!+\! 196217 z^4 \!-\! 734178 z^5 \!+\! 1145572 z^6 \!-\! 236240 z^7 \!-\! 21600 z^8 \!+\! 6432 z^9)+ d^3 (2264148- 3324172\, z  \\ \hiderel{}\!\!\!\!\!\!\!\!\!- 2560626 \, z^2 + 4208820\, z^3 - 2310448\, z^4 + 2819176\, z^5 - 2268746 z^6 \!+\! 267960 z^7  + 205232 z^8  \\ \hiderel{}\!\!\!\!\!\!\!\!\! \!-\! 57184 z^9 \!+\! 992 z^{10}) + d^2 (- 1057077 \!+\! 770046 z \!+\! 3165710 z^2 \!-\! 5024144 z^3 \!+\! 4741735 z^4  - 4632974 z^5  \\ \hiderel{}\!\!\!\!\!\!\!\!\! \!+\! 2404368 z^6 \!+\! 278752 z^7 \!-\! 609712 z^8 + 151328 z^9 + 2560 z^{10} - 512 z^{11}) - 2 d (- 133650  - 79020 z  \\ \hiderel{}\!\!\!\!\!\!\!\!\! \!+\! 999513 z^2 \!-\! 1814186 z^3 \!+\! 2177468 z^4 - 1806584 z^5 + 499333 z^6 + 468702 z^7  - 375688 z^8  + 58112 \,z^9  \\ \hiderel{}\!\!\!\!\!\!\!\!\! + 20944 \,z^{10} - 5088 z^{11} + 512 z^{12}) + 4 (\,-6561 - 23490 \,z + 134703 \,z^2 - 290208 \,z^3 + 381553 z^4  \\ \hiderel{}\!\!\!\!\!\!\!\!\! 
\!-\! 254922 z^5\!-\! 27091 z^6 \!+\! 160612 z^7 \!-\! 82092 z^8 \!-\! 4216 z^9 \!+\! 15520 z^{10} - 4320 z^{11} + 512 z^{12})\Big) \\ \hiderel{}\!\!\!\!\!\!\!\!\!
 + l^4 (z- 1) \Big(48 d^{10} + 6 d^9 (- 213  + 257 z) + d^8 (13865 	- 20391 z - 3434 z^2) + d^7 (- 80167 + 109448 z  \\ \hiderel{}\!\!\!\!\!\!\!\!\! + 46065 z^2 + 758 z^3) + d^6 (277079 - 301787 z - 261672 z^2 - 10010 z^3 + 2654 z^4) 	+ d^5 (- 604465  \\ \hiderel{}\!\!\!\!\!\!\!\!\! + 422090 z + 816557 z^2 + 71498 z^3 - 32848 z^4 - 1096 z^5) + d^4 (847481 - 156491 z - 1523076 z^2 \\ \hiderel{}\!\!\!\!\!\!\!\!\!  - 276702 z^3 + 139852 z^4 + 15200 z^5 - 816 z^6) -  d^3 (753973 + 386740 z - 1728795 z^2 - 591394 z^3  \\ \hiderel{}\!\!\!\!\!\!\!\!\! + 279236 z^4 + 68456 z^5 - 7952 z^6 + 288 z^7) + d^2 (404505 + 625291 z- 1153920 z^2 - 703870 z^3  \\ \hiderel{}\!\!\!\!\!\!\!\!\! + 277954 z^4 + 139184 z^5 - 21376 z^6 - 2080 z^7 + 992 z^8) + 6 (2133 + 14787 z - 8655 z^2  
- 18951 z^3  \\ \hiderel{}\!\!\!\!\!\!\!\!\! + 2190 z^4 + 8512 z^5 - 704 z^6 - 1392 z^7 + 224 z^8 + 128 z^9) -  d (115893 + 381684 z - 403191 z^2 \\ \hiderel{}\!\!\!\!\!\!\!\!\!  - 440606 z^3 + 123980 z^4  + 134112 z^5 - 19936 z^6 - 9344 z^7 + 2688 z^8 + 384 z^9)\Big) \sigma\Big)\,,
\end{dmath}

\begin{dmath}
c_1=\frac{4 l^2 ( d- 1 - 2 z)^2}{3\,\Xi \, (d- 1)} \Big(b_1 (d- 1)^2 \Big(192\, d^{10} - 12 d^9 (167 + 9\, z) + d^8 (\,8174 + 2276 \,z - 850 \,z^2)  \\ \hiderel{}\!\!\!\!\!\!\!\!\! -  d^7 (15125 + 16127 z - 7897 z^2  + 261 z^3) + d^6 (6543 + 55761 z - 29143 z^2 + 383 z^3 + 1016 z^4)  \\ \hiderel{}\!\!\!\!\!\!\!\!\! + 2 d^5 (11468 - 52082 z + 24603 z^2 + 4179 z^3 - 3953 z^4 + 329 z^5) 
- 4 (z- 1)^3 (- 9- 66 z - 481 z^2  \\ \hiderel{}\!\!\!\!\!\!\!\!\! + 2424 z^3 - 696 z^4 - 584 z^5 + 384 z^6) - 2 d^4 (21939 - 51140 z + 9653 z^2 + 24198 z^3 - 14750 z^4 \\ \hiderel{}\!\!\!\!\!\!\!\!\!  + 1320 z^5 + 524 z^6) + d^3 (32965 - 40005 z - 58437 z^2 + 116313 z^3 - 65592 z^4 + 4240 z^5 + 7020 z^6  \\ \hiderel{}\!\!\!\!\!\!\!\!\! - 1304 z^7) + d^2 (- 10955 - 7665 z + 86779 z^2 - 129523 z^3  + 70072 z^4 + 5084 z^5 	- 16680 z^6 + 3272 z^7  \\ \hiderel{}\!\!\!\!\!\!\!\!\!+ 192 z^8) + 2 d (594 + 3954 z - 18029 z^2 + 22111 z^3 - 471 z^4 - 17497 z^5 + 9594 z^6 + 1536 z^7  - 2304 z^8  \\ \hiderel{}\!\!\!\!\!\!\!\!\! + 512 z^9)\Big) -  l^4 (z-1) \Big(96 d^7 - 48 d^6 (13 + 11 z) + 2 d^5 (701 + 1596 z + 487 z^2) -  d^4 (973 + 7024 z  \\ \hiderel{}\!\!\!\!\!\!\!\!\! + 5317 z^2 + 606 z^3) + d^3 (- \!847 +6324 z +10917 z^2 +3006 z^3 -104 z^4) +d^2 (1693 -880 z -10235 z^2 \\ \hiderel{}\!\!\!\!\!\!\!\!\! -5566 z^3 +36 z^4 +168 z^5) +4 (36 +228 z +15 z^2 -593 z^3 +54 z^4 +84 z^5 -40 z^6) +d (-891 -1996 z  \\ \hiderel{}\!\!\!\!\!\!\!\!\! +3817 z^2 +5058 z^3+ 100 z^4 -424 z^5 +96 z^6)\Big) \sigma\Big)\,,
\end{dmath}
\begin{dmath}
c_2=\frac{2 l^2 (d- 1 - 2 z)}{\Xi (d- 1)} \Big(- b_1 (d- 1)^2 \Big(96 d^{11} - 124 d^{10} (7 + 5 z) + d^9 (1748 + 8728 z - 1020 z^2)  \\ \hiderel{}\!\!\!\!\!\!\!\!\! + d^8 (9153 - 51719 z + 9047 z^2  + 1119 z^3)  + 2 d^7 (- 30663 + 83293 z - 12584 z^2 - 8167 z^3 + 1169 z^4)  \\ \hiderel{}\!\!\!\!\!\!\!\!\! + d^6 (159487 - 309623\, z - 9071 \,z^2 + 98107\, z^3 - 22348 z^4 + 72 z^5)  
+ d^5 (- 226438 + 310048\, z  \\ \hiderel{}\!\!\!\!\!\!\!\!\! + 209684 z^2 - 313896 z^3 + 83778 z^4 + 4168 z^5 - 2496 z^6) -  d^4 (- 177839 + 93857 z + 521423 z^2   \\ \hiderel{}\!\!\!\!\!\!\!\!\! - 570561 z^3   + 136684 z^4+ 47248 z^5 - 21948 z^6 + 752 z^7) + 2 d^3 (- 31245 - 58491 z + 303532 z^2  \\ \hiderel{}\!\!\!\!\!\!\!\!\! - 277487 z^3 \!+\! 8835 z^4 + 95976 z^5 \!-\! 39536 z^6 \!+\! 1112 z^7  + 784 z^8) + d^2 (- 4979 + 118111 z - 323489 z^2  \\ \hiderel{}\!\!\!\!\!\!\!\!\! + 177297 z^3 + 270180 z^4 - 366208 z^5 + 127992 z^6 + 9176 z^7 - 9552 z^8 + 896 z^9) + 4 (- 378 - 1001 z \\ \hiderel{}\!\!\!\!\!\!\!\!\!  + 8250\, z^2 - 26583\, z^3 + 41305 \,z^4 - 25338 \,z^5 - 3141 z^6 + 9802 z^7 - 2460 z^8 - 840 z^9 + 384 z^{10})  \\ \hiderel{}\!\!\!\!\!\!\!\!\! - 2 d (- 4645 + 13334 \,z - 11392 \,z^2 - 69538 \,z^3 + 187549 \,z^4 - 162044 z^5 + 34752 z^6 + 20800 z^7  \\ \hiderel{}\!\!\!\!\!\!\!\!\! - 9072 z^8 - 256 z^9 + 512 z^{10})\Big) + l^4 (z- 1) \Big(- 495 + 48 d^8 - 4272 z - 3303 z^2 + 8030 z^3 + 2248 z^4  \\ \hiderel{}\!\!\!\!\!\!\!\!\!- 1584 z^5 + 16 z^6 + 224 z^7 - 6 d^7 (41 + 91 z) + d^6 (- 181 + 3891 z + 1618 z^2) -  d^5 (- 3910 + 9625 z  \\ \hiderel{}\!\!\!\!\!\!\!\!\! + 11529 z^2 + 1692 z^3)+ d^4 (- 11339 + 6684 z + 31675 z^2 + 11036 z^3 \!+\! 392 z^4) + d^2 (- 11193\! -\! 24191 z \\ \hiderel{}\!\!\!\!\!\!\!\!\!  + 23170 z^2 + 38126 z^3 + 4240 z^4 - 1112 z^5) + 2 d^3 (7780 + 5571 z - 20590 z^2 - 14379 z^3 - 1018 z^4  \\ \hiderel{}\!\!\!\!\!\!\!\!\! + 136 z^5) -  d (- 3936 - 16917 z + 707 z^2 + 26502 z^3 + 4444 z^4 - 2024 z^5 + 96 z^6 + 128 z^7)\Big) \sigma\Big)\,,
\end{dmath}
\begin{dmath}
c_3=\frac{l^2}{3\, \Xi \, (d-1)} \Big(2 b_1 (d-1)^2 \Big(96 d^{11} (2 + z) - 4 d^{10} (542 + 647 z + 239 z^2) + 2 d^9 (4137 + 13375 z \\ \hiderel{}\!\!\!\!\!\!\!\!\! + 5821 z^2 \!+\! 811 z^3) + d^8 (-575 \!-\! 148059 z \!-\! 67177 z^2 \!-\! 14765 z^3 + 272 z^4) + d^7 (-104799 + 495534 z \\ \hiderel{}\!\!\!\!\!\!\!\!\! + 254339 z^2 + 51655 z^3 - 7448 z^4 + 1055 z^5) -  d^6 (-422579 + 1048799 z + 705765 z^2 + 92573 z^3 \\ \hiderel{}\!\!\!\!\!\!\!\!\! - 66154 z^4 + 6916 z^5 + 4984 z^6) + d^5 (-883077 + 1391766 z + 1439961 z^2 + 131129 z^3 - 273580 z^4 \\ \hiderel{}\!\!\!\!\!\!\!\!\!- 11363 z^5 + 51308 z^6 + 416 z^7) + d^4 (1130881 - 1050285 z - 2087619 z^2 - 240425 z^3 + 554776 z^4 \\ \hiderel{}\!\!\!\!\!\!\!\!\!+ 208454 z^5 - 206350 z^6 - 11640 z^7 + 3680 z^8) + d^3 (-909813 + 246562 z + 2092313 z^2 + 299465 z^3 \\ \hiderel{}\!\!\!\!\!\!\!\!\!- 420636 z^4 - 702839 z^5 + 383884 z^6 + 92536 z^7 - 30160 z^8 + 272 z^9) + d^2 (442189 + 266887 z\\ \hiderel{}\!\!\!\!\!\!\!\!\!  - 1429703 z^2 - 22723 z^3 - 266418 z^4 + 1043240 z^5 - 257136 z^6 - 275472 z^7 + 75120 z^8 + 6240 z^9 \\ \hiderel{}\!\!\!\!\!\!\!\!\! - 2560 z^{10})\! -\! 2 (-5787 \!-\! 30438 z + 75230 z^2 \!-\! 86563 z^3 + 125480 z^4 - 49747 z^5 - 85555 z^6 + 67076 z^7\\ \hiderel{}\!\!\!\!\!\!\!\!\! + 7760 z^8 \!-\! 14560 z^9 \!+\! 1520 z^{10} \!+\! 768 z^{11}) + d (-115257 \!-\! 238740 z \!+\! 641505 z^2 \!-\! 282383 z^3 + 604464 z^4 \\ \hiderel{}\!\!\!\!\!\!\!\!\! - 656757 z^5 - 118152 z^6 + 335960 z^7 - 49600 z^8 - 29968 z^9 + 6400 z^{10} + 1024 z^{11})\Big)\\ \hiderel{}\!\!\!\!\!\!\!\!\! 
 + l^4 (z-1) \Big(-5553 \!+\! 48 d^9 \!-\! 44754 z \!-\! 25173 z^2 \!+\! 48772 z^3 \!+\! 38020 z^4 \!-\! 6080 z^5 - 4912 z^6+ 1856 z^7 \\ \hiderel{}\!\!\!\!\!\!\!\!\! + 704 z^8 + 6 d^8 (-221 + 233 z) + d^7 (12479 - 12963 z - 5732 z^2) + d^6 (-58800 + 41258 z + 54524 z^2 \\ \hiderel{}\!\!\!\!\!\!\!\!\!+ 8306 z^3) -  d^5 (-159342 + 32779 z + 204587 z^2 + 71256 z^3 + 4776 z^4) + d^4 (-262123 - 111044 z \\ \hiderel{}\!\!\!\!\!\!\!\!\! + 385791 z^2 + 244444 z^3 + 33276 z^4 + 368 z^5) + d^3 (263275 + 325003 z - 374458 z^2 - 433644 z^3 \\ \hiderel{}\!\!\!\!\!\!\!\!\! - 94216 z^4 + 800 z^5 + 64 z^6) + 2 d^2 (-77539 - 184165 z + 75453 z^2 + 212799 z^3 + 68792 z^4 - 3504 z^5 \\ \hiderel{}\!\!\!\!\!\!\!\!\! - 528 z^6 + 336 z^7) + d (47736 + 202211 z + 18057 z^2 - 222796 z^3 - 107296 z^4 + 11824 z^5 + 4304 z^6 \\ \hiderel{}\!\!\!\!\!\!\!\!\! - 2496 z^7 - 384 z^8)\Big) \sigma\Big)
\end{dmath}
\begin{dmath}
\L_0=\frac{(d- 1)^2(z-1)(d-2)}{6l^4\, \Xi} \Big(b_1 (d-1 )^2 \Big(96\, d^{13} - 4\, d^{12} (301 +167\, z) + 2\, d^{11} (2321 + 4446\, z \\ \hiderel{}\!\!\!\!\!\!\!\!\! + 649 z^2) + d^{10} (4494 - 37194 z - 32138 z^2 + 566 z^3)  -  d^9 (91761+ 46637 z - 332843 z^2 + 19043 z^3 \\ \hiderel{}\!\!\!\!\!\!\!\!\!+ 4474 z^4) + 2 d^8 (142628 + 552660 z - 978547 z^2 + 116805 z^3 + 19033 z^4 + 325 z^5)  + d^7 (- 209009\\ \hiderel{}\!\!\!\!\!\!\!\!\!
- 5243745 z \!+\! 7366309 z^2 \!-\! 1506913 z^3 \!-\! 54008 z^4 \!-\! 29014 z^5 \!+\! 10412 z^6) \!-\! 4 d^6 (242497 - 3500960 z \\ \hiderel{}\!\!\!\!\!\!\!\!\!  + 4677846 z^2   - 1425308 z^3
+ 74845 z^4 - 22604 z^5 + 15135 z^6 + 577 z^7) + d^5 (3326409 - 23996307 z \\ \hiderel{}\!\!\!\!\!\!\!\!\!+ 32437993 \,z^2 - 12680361 \,z^3 + 32802 z^4  + 1250264\, z^5- 354100 \,z^6 + 111284 \,z^7 - 12352 \,z^8) \\ \hiderel{}\!\!\!\!\!\!\!\!\! 
- 2 d^4 (2549650 \!-\! 13499190 z \!+\! 18601679 z^2 \!-\! 7152261 z^3 - 4002317 z^4  + 5158255 z^5- 2070480 z^6 \\ \hiderel{}\!\!\!\!\!\!\!\!\!
+ 492404 z^7 - 64572 z^8 + 2096 z^9) + 24 (z- 1 )^2 (- 2673 \!-\! 6837 z \!+\! 59030 z^2 \!-\! 194638 z^3 \!+\! 184895 z^4 \\ \hiderel{}\!\!\!\!\!\!\!\!\!  - 51097 z^5 - 20212 z^6 
+ 20156 z^7 - 6232 z^8 - 16 z^9 + 128 z^{10}) - 4 d (z- 1 )^2 (- 156789 + 27327 z \\ \hiderel{}\!\!\!\!\!\!\!\!\! + 869344 z^2 - 3597020 z^3 + 3281227 z^4 - 1011643 z^5 - 63478 z^6 + 146288 z^7 - 50544 z^8 + 864 z^9 \\ \hiderel{}\!\!\!\!\!\!\!\!\!  + 384 z^{10}) + d^3 (4507267 - 19319383 z + 25346473 z^2 - 450439 z^3  - 31360644 z^4 + 33041610 z^5\\ \hiderel{}\!\!\!\!\!\!\!\!\!
- 14844876 z^6  + 3466128 z^7 - 362680 z^8 - 33696 z^9 + 13696 z^{10}) + 2 d^2 (- 1159953 + 3963453 z \\ \hiderel{}\!\!\!\!\!\!\!\!\! - 3307361 z^2 - 9615401 z^3 
+ 26992568 z^4   - 27014098 z^5 + 12737522 z^6 - 2689998 z^7 - 52956 z^8\\ \hiderel{}\!\!\!\!\!\!\!\!\! + 192432 z^9 - 47072 z^{10} + 864 z^{11})\Big) + l^4 \big(- 48 d^{11} - 6 d^{10} (- 229 + 305 z) + d^9 (- 16133 + 29463 z \\ \hiderel{}\!\!\!\!\!\!\!\!\! - 2026 z^2) + d^8 (103997 - 211556 z + 33497 z^2 - 1754 z^3) + d^7 (- 414271  + 888009 z - 213628 z^2 \\ \hiderel{}\!\!\!\!\!\!\!\!\! - 9136 z^3 + 14346 z^4) + d^6 (1079933 - 2392440 z + 665429 z^2 + 252988 z^3 - 157986 z^4 - 10316 z^5)\\ \hiderel{}\!\!\!\!\!\!\!\!\! + d^5 (- 1888171 \!+\! 4263303 z 
\!-\! 938908 z^2 \!-\! 1474234 z^3 \!+\! 694666 z^4 \!+\! 104844 z^5 \!+\! 668 z^6) \!+\! d^4 (2215571 \\ \hiderel{}\!\!\!\!\!\!\!\!\! - 4973808 z - 98209 z^2 + 4201870 z^3 \!-\! 1567856 z^4  \!-\! 441772 z^5 \!-\! 3404 z^6 + 1056 z^7) + d^3 (- 1703657 \\ \hiderel{}\!\!\!\!\!\!\!\!\! + 3591603 z + 2358120 z^2 - 6831716 z^3 + 2037678 z^4  + 869616 z^5 + 60772 z^6 
- 20904 z^7 + 3456 z^8) \\ \hiderel{}\!\!\!\!\!\!\!\!\! + 12 (z-1)^2 (1755 + 7281 z - 37941 z^2 - 1305 z^3 - 9142 z^4 - 1840 z^5 + 3464 z^6 - 704 z^7  - 512 z^8 \\ \hiderel{}\!\!\!\!\!\!\!\!\! + 64 z^9)\! -\!  d^2 (- 809505 \!+\! 1374322 z \!+\! 3511149 z^2 - 6627432 z^3 + 1817538 z^4 + 654796 z^5 + 243420 z^6 \\ \hiderel{}\!\!\!\!\!\!\!\!\! - 48112 z^7 - 8464 z^8 + 6336 z^9) + 2 d (- 104580 + 68163 z + 1156005 z^2 - 1866385 z^3 + 658415 z^4 \\ \hiderel{}\!\!\!\!\!\!\!\!\! - 36670 z^5 + 137128 z^6 + 15724 z^7 - 27752 z^8 + 7440 z^9 + 288 z^{10})\Big) \sigma\Big)\,,
\end{dmath}
\begin{dmath}\label{L0BHLS}
\Xi=(d -  2) (d- 1) l^2 (z- 1) \Big(48\, d^{10} + 6\, d^9 (- 205 + 281 \,z) + d^8 (13283 - 23739\, z - 848\, z^2) \\ \hiderel{}\!\!\!\!\!\!\!\!\! - 2 d^7 (38719 - 70645 z - 6983 z^2 + 1001 z^3) + d^6 (271048 - 460549 z - 106277 z^2 + 32506 z^3 \\ \hiderel{}\!\!\!\!\!\!\!\!\!- 2048 z^4) + d^5 (- 599393 + 881958 z + 453483 z^2 - 187002 z^3 + 7594 z^4 + 5752 z^5) + d^4 (851281 \\ \hiderel{}\!\!\!\!\!\!\!\!\! - 963709 z - 1153994 z^2 + 518884 z^3 + 35970 z^4 - 48024 z^5 - 2576 z^6) - 2 d^3 (382936 - 235875 z \\ \hiderel{}\!\!\!\!\!\!\!\!\! - 891966 z^2 + 379545 z^3 + 111998 z^4 - 70298 z^5 - 10128 z^6 + 512 z^7) + d^2 (414374 + 97301 z \\ \hiderel{}\!\!\!\!\!\!\!\!\!- 1643993 z^2 + 587670 z^3 + 399372 z^4 - 160156 z^5 - 65600 z^6 
+ 5104 z^7 + 960 z^8) + d (- 119187 \\ \hiderel{}\!\!\!\!\!\!\!\!\! - 216572 z + 834955 z^2 - 231586 z^3 - 273774 z^4 + 43412 z^5 + 82736 z^6 - 592 z^7 - 6624 z^8 + 1280 z^9) \\ \hiderel{}\!\!\!\!\!\!\!\!\! + 2 (6543 + 35292 z - 90804 z^2 + 21142 z^3 + 28009 z^4 + 6090 z^5 - 10736 z^6 - 6656 z^7 + 3120 z^8 \\ \hiderel{}\!\!\!\!\!\!\!\!\! + 736 z^9 - 512 z^{10})\Big)\,,
\end{dmath}
\end{dgroup}
\pagebreak
\section{Counter term coefficients}
The coefficients of counter-terms in equation (\ref{SCC}) are as follows

\begin{dgroup}
	\begin{dmath}
		\\ \hiderel{} \!\!\!\! \a_1=(- b_1 (d-1)^3 (-395016343175136 + 2348550175021544 d - 3316571455059268 d^2  \\ \hiderel{} - 9312466632690362 d^3 + 41832587177120467 d^4 - 59950661474513668 d^5  \\ \hiderel{} + 4652385337180473 d^6  + 128519961747254140 d^7 - 255642279986199850 d^8 \\ \hiderel{} + 284182844961412260 d^9 - 209960379268229286 d^{10} + 106564661277557906 d^{11} \\ \hiderel{} - 36670939831292789 d^{12} + 8074817985963160 d^{13} - 943970979676719 d^{14} \\ \hiderel{} + 10294748426892 d^{15} + 7149242979468 d^{16})  + 2 (-2507607366432  + 9370485047140 d \\ \hiderel{} - 15322963757344 d^2 + 29194387485479 d^3  - 34599990417180 d^4 - 101018121331169 d^5\\ \hiderel{}+ 569000982666250 d^6 - 1443232620917570 d^7 + 2383139360258160 d^8 \\ \hiderel{}- 2637960123419686 d^9 + 1920556277058316 d^{10} - 892060727009845 d^{11} \\ \hiderel{} + 251884406261660 d^{12} - 38888689855581 d^{13}  + 1851979430970 d^{14} \\ \hiderel{}  + 344868283728 d^{15}) l^4 \sigma )/\U\,,
	\end{dmath}

	\begin{dmath}
		\\ \hiderel{} \!\!\!\! \a_2= 4 l^2 \big(2 b_1 (d-1)^3 (-76757505990096 + 452153505555884 d - 624980807053248 d^2 \\ \hiderel{} - 1809192261855057 d^3 + 7952875768988962 d^4 - 11226865296348523 d^5 \\ \hiderel{} + 740755958471253 d^6 + 23887391502121540 d^7 - 46975129179219225 d^8 \\ \hiderel{} + 51879608281499860 d^9  - 38210420906572496 d^{10} + 19391626011252941 d^{11} \\ \hiderel{} - 6690472887584954 d^{12}  + 1480665214857635 d^{13} - 174392189027309 d^{14} \\ \hiderel{} + 1988579461512 d^{15} + 1327387196073 d^{16}) + (1990194128808 - 7386025312660 d  \\ \hiderel{} + 11742213076986 d^2 - 21594332747301 d^3 + 23694835045670 d^4 + 83844566027911 d^5  \\ \hiderel{} - 437291091342750 d^6 + 1076981245232830 d^7  - 1754567637812540 d^8  \\ \hiderel{} + 1934727144371834 d^9  - 1409181677959554 d^{10}  + 655519789230055 d^{11}  \\ \hiderel{} - 185368649436290 d^{12} + 28733073480339 d^{13} - 1405718397930 d^{14}  \\ \hiderel{}- 254604527232 d^{15}) l^4 \sigma\big)/(d-1)\U\,,
	\end{dmath}
	\begin{dmath}
		\\ \hiderel{} \!\!\!\! \a_3= 4 l^4 \big(- b_1 (d-1)^3 (1408481241557088 - 7962097900340552 d + 8702328896412548 d^2  \\ \hiderel{} + 40621533255118130 d^3 - 144622783539983159 d^4 + 152594357275463212 d^5  \\ \hiderel{} + 118911670846630178 d^6 - 560396683643759318 d^7 + 764990748437182203 d^8  \\ \hiderel{} - 490234816648922540 d^9 - 27828284889708812 d^{10} + 361280304561738662 d^{11}  \\ \hiderel{} - 362359563639109409 d^{12} + 202400102173512036 d^{13} - 71325089681437902 d^{14}  \\ \hiderel{} + 15535905541976574 d^{15} - 1742569596681603 d^{16} + 10872378426212 d^{17}  \\ \hiderel{} + 13031320359348 d^{18}) + 2 (9079919253456 - 31647743507020 d + 41985077042700 d^2  \\ \hiderel{} - 71966579675217 d^3 + 59972441260106 d^4 + 475565980123096 d^5 - 2033918977126280 d^6  \\ \hiderel{} + 4370244951580351 d^7 - 5843504482948030 d^8 + 4135073914235468 d^9  \\ \hiderel{} + 309968734691532 d^{10}  - 3588169908956611 d^{11} + 3470610662200846 d^{12}  \\ \hiderel{} - 1717595387287272 d^{13} + 484209291110800 d^{14} - 72910129367115 d^{15}  \\ \hiderel{} + 3042549337270 d^{16} + 609885548208 d^{17}) l^4 \sigma\big)/9 (d-1)^4 \U\,,
	\end{dmath}
	\begin{dmath}
		\\ \hiderel{} \!\!\!\! \a_4= 8 (d-2) l^4 \big(b_1 (d-1)^3 (-528808080422496 + 3130553990186984 d  \\ \hiderel{}- 4379020454475448 d^2   - 12459226694726482 d^3 + 55439333803203762 d^4  \\ \hiderel{} - 78953331948719473 d^5 + 5806789128993578 d^6 + 168554151710199165 d^7  \\ \hiderel{} - 333816422809221600 d^8 + 370185524282685110 d^9 - 273220870869987396 d^{10}  \\ \hiderel{} + 138699652880065216 d^{11} - 47789659274303054 d^{12} + 10546137780003635 d^{13}  \\ \hiderel{} - 1236609482027534 d^{14} + 13691674656437 d^{15} + 9391169673148 d^{16})  \\ \hiderel{} + 2 (3384337452252 - 12612711742540 d + 20346088117159 d^2 - 38096175355894 d^3  \\ \hiderel{} + 43891199458605 d^4 + 138169028535484 d^5 - 755344603837500 d^6 + 1895200175897020 d^7  \\ \hiderel{} - 3114206365072010 d^8 + 3442492641342796 d^9 - 2506736133351601 d^{10}  \\ \hiderel{} + 1164981189763170 d^{11} - 329152871207135 d^{12} + 50916198362516 d^{13}  \\ \hiderel{} - 2456245454170 d^{14} - 450654150008 d^{15}) l^4 \sigma\big)/9 (d-1)^3\U\,,
	\end{dmath}
	\begin{dmath}
		\\ \hiderel{} \!\!\!\! \a_5= - 4 l^6 \big(- b_1 (d-1)^3 (359778834028416 - 4953309366088160 d + 21513244670246864 d^2 \\ \hiderel{} - 25733295796545048 d^3 - 88916513146380432 d^4 + 392638707702220032 d^5 \\ \hiderel{}- 618350454473661033 d^6 + 191645008663521032 d^7 + 1108939951725441017 d^8\\ \hiderel{} - 2628098535293123038 d^9 + 3328476896217706846 d^{10} - 2826396397716689052 d^{11}\\ \hiderel{} + 1683301856093304062 d^{12} - 693105132653871760 d^{13} + 181474277097201271 d^{14}\\ \hiderel{} - 20572142542150452 d^{15} - 3854293317416223 d^{16} + 1889784234492878 d^{17} \\ \hiderel{}- 262274701598268 d^{18} + 2960158017632 d^{19} + 1786518075960 d^{20}) + 4 (1138565977896 \\ \hiderel{} - 13135644442796 d + 45147119557134 d^2 - 84986822962523 d^3 + 146870086724791 d^4 \\ \hiderel{}- 145372529469655 d^5 - 498268517295789 d^6 + 2798356123102759 d^7 \\ \hiderel{}- 7314741050355708 d^8 + 12688604682534093 d^9 - 15370507537963346 d^{10} \\ \hiderel{}+ 12921839882975879 d^{11} - 7328568420878773 d^{12} + 2630599453118935 d^{13}\\ \hiderel{} - 488404739538533 d^{14} - 11229759183123 d^{15} + 27825947094802 d^{16} \\ \hiderel{}- 5713942549473 d^{17} + 262503065366 d^{18} + 42373630080 d^{19}) l^4 \sigma\big)/9 (d-3) (d-1)^6 \U\,,
	\end{dmath}
	\begin{dmath}
		\\ \hiderel{} \!\!\!\! \a_6=- 2 (d-2) l^6 \big(b_1 (d-1)^3 (902871864042048 - 2611066522935696 d  \\ \hiderel{}- 10000441381807568 d^2+ 51596855614148496 d^3 - 41200165261664328 d^4 \\ \hiderel{}- 181595179962421999 d^5 + 533465796657080410 d^6 - 513376878534607377 d^7 \\ \hiderel{}- 279078582910266796 d^8 + 1491989115301771590 d^9 - 2247285100078274912 d^{10} \\ \hiderel{}+ 2066673831684584018 d^{11} - 1294190663758530772 d^{12} + 563666594089214817 d^{13}\\ \hiderel{} - 167708947843161610 d^{14} + 31807182182030383 d^{15} - 3067722379531768 d^{16}\\ \hiderel{} - 15879451896872 d^{17} + 22671972030304 d^{18}) - 4 (2899692106488 - 2065011520884 d \\ \hiderel{}- 19172437911422 d^2 + 34519082576341 d^3 - 83448423628318 d^4 + 277332351302717 d^5 \\ \hiderel{}- 353449049883778 d^6 - 467932776789536 d^7 + 3107362543650320 d^8 \\ \hiderel{}- 7360760025239886 d^9 + 10493493512297994 d^{10} - 9626241270064687 d^{11} \\ \hiderel{}+ 5748085802510894 d^{12} - 2210046214383931 d^{13} + 525872565846518 d^{14} \\ \hiderel{}- 68068023018246 d^{15} + 1803501433576 d^{16} + 540861289792 d^{17}) l^4 \sigma\big)/9 (d-3 ) (d-1)^5 \U\,,
	\end{dmath}
	\begin{dmath}
		\\ \hiderel{} \!\!\!\! \a_7= 2 (d-2)^2 l^6 \big(b_1 (d-1)^3 (-786581801966112 + 4658820413439448 d  \\ \hiderel{} - 6529543654029956 d^2 - 18497282599374254 d^3 + 82484042277274539 d^4  \\ \hiderel{}- 117690486380092256 d^5 + 9149225718692941 d^6 + 250315027315438380 d^7  \\ \hiderel{}- 496717867127149950 d^8 + 551374708000114420 d^9 - 407208903035254662 d^{10}  \\ \hiderel{}+ 206801753285171702 d^{11} - 71269711008561613 d^{12} + 15728365006707220 d^{13}  \\ \hiderel{}- 1844046118745923 d^{14} + 20374203626764 d^{15} + 14008411441856 d^{16})  \\ \hiderel{} - 4 (-2512500801972 + 9353193417940 d - 15028567971649 d^2 + 28141140546484 d^3  \\ \hiderel{}- 32887608597405 d^4 - 100908676070974 d^5 + 558339673947000 d^6 - 1407114644223220 d^7  \\ \hiderel{}+ 2317072816104110 d^8 - 2563402773401356 d^9 + 1866929648661211 d^{10}  \\ \hiderel{}- 867565615650120 d^{11} + 245083387747235 d^{12} - 37904656710026 d^{13}  \\ \hiderel{}+ 1827835429870 d^{14} + 335120714888 d^{15}) l^4 \sigma\big)/9 (d-3 ) (d-1 )^4 \U \,,
	\end{dmath}
	\begin{dmath}
		\\ \hiderel{} \!\!\!\! \U= 1000 (-4362 + 3115 d + 17033 d^2 - 32164 d^3 + 21034 d^4 - 5675 d^5 + 631 d^6 + 68 d^7) (-120984   \\ \hiderel{} + 504826 d  - 662169 d^2 + 182970 d^3 + 106438 d^4 + 252706 d^5 - 415237 d^6 + 154042 d^7) l^5\,.
	\end{dmath}
\end{dgroup}
\section*{Acknowledgment}
This work is supported by Ferdowsi University of Mashhad under the grant 3/39197
(1394/12/26).
\providecommand{\href}[2]{#2}\begingroup\raggedright

\endgroup

\begin{thebibliography}{99}


\bibitem{Stelle:1976gc} 
K.~S.~Stelle,
``Renormalization of Higher Derivative Quantum Gravity,''
Phys.\ Rev.\ D {\bf 16}, 953 (1977).

\bibitem{Zwiebach:1985uq} 
B.~Zwiebach,
``Curvature Squared Terms and String Theories,''
Phys.\ Lett.\ B {\bf 156}, 315 (1985).

\bibitem{Metsaev:1986yb} 
R.~R.~Metsaev and A.~A.~Tseytlin,
``Curvature Cubed Terms in String Theory Effective Actions,''
Phys.\ Lett.\ B {\bf 185}, 52 (1987).

\bibitem{Buchel:2009sk} 
A.~Buchel, J.~Escobedo, R.~C.~Myers, M.~F.~Paulos, A.~Sinha and M.~Smolkin,
``Holographic GB gravity in arbitrary dimensions,''
JHEP {\bf 1003}, 111 (2010)
[arXiv:0911.4257 [hep-th]].

\bibitem{Myers:2010tj} 
R.~C.~Myers and A.~Sinha,
``Holographic c-theorems in arbitrary dimensions,''
JHEP {\bf 1101}, 125 (2011)
[arXiv:1011.5819 [hep-th]].

\bibitem{Bergshoeff:2012sc} 
E.~A.~Bergshoeff, S.~de Haan, W.~Merbis, M.~Porrati and J.~Rosseel,
``Unitary Truncations and Critical Gravity: a Toy Model,''
JHEP {\bf 1204}, 134 (2012)
[arXiv:1201.0449 [hep-th]].


\bibitem{Bergshoeff:2012ev} 
E.~A.~Bergshoeff, S.~de Haan, W.~Merbis, J.~Rosseel and T.~Zojer,
``On Three-Dimensional Tricritical Gravity,''
Phys.\ Rev.\ D {\bf 86}, 064037 (2012)
[arXiv:1206.3089 [hep-th]].

\bibitem{Nutma:2012ss} 
T.~Nutma,
``Polycritical Gravities,''
Phys.\ Rev.\ D {\bf 85}, 124040 (2012)
[arXiv:1203.5338 [hep-th]].

\bibitem{Kleinschmidt:2012rs} 
A.~Kleinschmidt, T.~Nutma and A.~Virmani,
``On unitary subsectors of polycritical gravities,''
Gen.\ Rel.\ Grav.\  {\bf 45}, 727 (2013)
[arXiv:1206.7095 [hep-th]].


\bibitem{Lu:2013hx} 
H.~Lü, Y.~Pang and C.~N.~Pope,
``Black Holes in Six-dimensional Conformal Gravity,''
Phys.\ Rev.\ D {\bf 87}, no. 10, 104013 (2013)
[arXiv:1301.7083 [hep-th]].

\bibitem{Apolo:2012vv} 
L.~Apolo and M.~Porrati,
``Nonlinear Dynamics of Parity-Even Tricritical Gravity in Three and Four Dimensions,''
JHEP {\bf 1208}, 051 (2012)
[arXiv:1206.5231 [hep-th]].


\bibitem{Deser:1982ac}
  S.~Deser, R.~Jackiw and S.~Templeton,
{\it Topologically massive gauge theories},
  Annals Phys.\  {\bf 140} (1982) 372.
  [Erratum-ibid.\  {\bf 185}, 406.1988\ APNYA,281,409
(1988\ APNYA,281,409-449.2000)].

\bibitem{Li:2008dq} 
  W.~Li, W.~Song and A.~Strominger,
  ``Chiral Gravity in Three Dimensions,''
  JHEP {\bf 0804}, 082 (2008)
  [arXiv:0801.4566 [hep-th]].


\bibitem{Bergshoeff:2009hq} 
E.~A.~Bergshoeff, O.~Hohm and P.~K.~Townsend,
``Massive Gravity in Three Dimensions,''
Phys.\ Rev.\ Lett.\  {\bf 102}, 201301 (2009)
[arXiv:0901.1766 [hep-th]].

\bibitem{Bergshoeff:2009aq} 
E.~A.~Bergshoeff, O.~Hohm and P.~K.~Townsend,
``More on Massive 3D Gravity,''
Phys.\ Rev.\ D {\bf 79}, 124042 (2009)
[arXiv:0905.1259 [hep-th]].


\bibitem{Afshar:2014ffa} 
H.~R.~Afshar, E.~A.~Bergshoeff and W.~Merbis,
``Extended massive gravity in three dimensions,''
JHEP {\bf 1408}, 115 (2014)
[arXiv:1405.6213 [hep-th]].

\bibitem{Sinha:2010ai} 
A.~Sinha,
``On the new massive gravity and AdS/CFT,''
JHEP {\bf 1006}, 061 (2010)
[arXiv:1003.0683 [hep-th]].


\bibitem{Paulos:2010ke} 
M.~F.~Paulos,
``New massive gravity extended with an arbitrary number of curvature corrections,''
Phys.\ Rev.\ D {\bf 82}, 084042 (2010)
[arXiv:1005.1646 [hep-th]].

\bibitem{Ghodsi:2010gk} 
A.~Ghodsi and M.~Moghadassi,
``Charged Black Holes in New Massive Gravity,''
Phys.\ Lett.\ B {\bf 695}, 359 (2011)
[arXiv:1007.4323 [hep-th]].

\bibitem{Ghodsi:2011ua} 
A.~Ghodsi and D.~M.~Yekta,
``On Asymptotically AdS-Like Solutions of Three Dimensional Massive Gravity,''
JHEP {\bf 1206}, 131 (2012)
[arXiv:1112.5402 [hep-th]].

\bibitem{Ghodsi:2012fg} 
A.~Ghodsi and D.~M.~Yekta,
``Stability of vacua in New Massive Gravity in different gauges,''
JHEP {\bf 1308}, 095 (2013)
[arXiv:1212.6876 [hep-th]].

\bibitem{Myers:2010ru} 
  R.~C.~Myers and B.~Robinson,
  ``Black Holes in Quasi-topological Gravity,''
  JHEP {\bf 1008}, 067 (2010)
  [arXiv:1003.5357 [gr-qc]].

\bibitem{Lu:2011ks} 
  H.~Lu, Y.~Pang and C.~N.~Pope,
  ``Conformal Gravity and Extensions of Critical Gravity,''
  Phys.\ Rev.\ D {\bf 84}, 064001 (2011)
  [arXiv:1106.4657 [hep-th]].


\bibitem{Gullu:2009vy} 
I.~Gullu and B.~Tekin,
``Massive Higher Derivative Gravity in D-dimensional Anti-de Sitter Spacetimes,''
Phys.\ Rev.\ D {\bf 80}, 064033 (2009)
[arXiv:0906.0102 [hep-th]].

\bibitem{Oliva:2010zd} 
J.~Oliva and S.~Ray,
``Classification of Six Derivative Lagrangians of Gravity and Static Spherically Symmetric Solutions,''
Phys.\ Rev.\ D {\bf 82}, 124030 (2010)
[arXiv:1004.0737 [gr-qc]].

\bibitem{Sisman:2011gz} 
T.~C.~Sisman, I.~Gullu and B.~Tekin,
``All unitary cubic curvature gravities in D dimensions,''
Class.\ Quant.\ Grav.\  {\bf 28}, 195004 (2011)
[arXiv:1103.2307 [hep-th]].

\bibitem{Bueno:2016xff} 
P.~Bueno and P.~A.~Cano,
``Einsteinian cubic gravity,''
Phys.\ Rev.\ D {\bf 94}, no. 10, 104005 (2016)
[arXiv:1607.06463 [hep-th]].


\bibitem{Bueno:2016ypa} 
P.~Bueno, P.~A.~Cano, V.~S.~Min and M.~R.~Visser,
``Aspects of general higher-order gravities,''
Phys.\ Rev.\ D {\bf 95}, no. 4, 044010 (2017)
[arXiv:1610.08519 [hep-th]].

\bibitem{Abbott:1981ff} 
L.~F.~Abbott and S.~Deser,
``Stability of Gravity with a Cosmological Constant,''
Nucl.\ Phys.\ B {\bf 195}, 76 (1982).

\bibitem{Deser:2002rt} 
S.~Deser and B.~Tekin,
``Gravitational energy in quadratic curvature gravities,''
Phys.\ Rev.\ Lett.\  {\bf 89}, 101101 (2002)
[hep-th/0205318].

\bibitem{Hohm:2010jc} 
O.~Hohm and E.~Tonni,
``A boundary stress tensor for higher-derivative gravity in AdS and Lifshitz backgrounds,''
JHEP {\bf 1004}, 093 (2010)
[arXiv:1001.3598 [hep-th]].


\bibitem{Deser:2011xc} 
S.~Deser, H.~Liu, H.~Lu, C.~N.~Pope, T.~C.~Sisman and B.~Tekin,
``Critical Points of D-Dimensional Extended Gravities,''
Phys.\ Rev.\ D {\bf 83}, 061502 (2011)
[arXiv:1101.4009 [hep-th]].



\bibitem{Deser:2002jk} 
S.~Deser and B.~Tekin,
``Energy in generic higher curvature gravity theories,''
Phys.\ Rev.\ D {\bf 67}, 084009 (2003)
[hep-th/0212292].


\bibitem{Bergshoeff:2011ri} 
E.~A.~Bergshoeff, O.~Hohm, J.~Rosseel and P.~K.~Townsend,
``Modes of Log Gravity,''
Phys.\ Rev.\ D {\bf 83}, 104038 (2011)
[arXiv:1102.4091 [hep-th]].

\bibitem{Dyer:2008hb} 
E.~Dyer and K.~Hinterbichler,
``Boundary Terms, Variational Principles and Higher Derivative Modified Gravity,''
Phys.\ Rev.\ D {\bf 79}, 024028 (2009)
[arXiv:0809.4033 [gr-qc]].

\bibitem{Balcerzak:2008bg} 
A.~Balcerzak and M.~P.~Dabrowski,
``Gibbons-Hawking Boundary Terms and Junction Conditions for Higher-Order Brane Gravity Models,''
JCAP {\bf 0901}, 018 (2009)
[arXiv:0804.0855 [hep-th]].

\bibitem{Jacobson:1993vj} 
T.~Jacobson, G.~Kang and R.~C.~Myers,
``On black hole entropy,''
Phys.\ Rev.\ D {\bf 49}, 6587 (1994)
[gr-qc/9312023].

\bibitem{Kachru:2008yh} 
S.~Kachru, X.~Liu and M.~Mulligan,
``Gravity duals of Lifshitz-like fixed points,''
Phys.\ Rev.\ D {\bf 78}, 106005 (2008)
[arXiv:0808.1725 [hep-th]].

\bibitem{Mann:2009yx} 
  R.~B.~Mann,
  ``Lifshitz Topological Black Holes,''
  JHEP {\bf 0906}, 075 (2009)
  [arXiv:0905.1136 [hep-th]].

\bibitem{Bertoldi:2009vn} 
  G.~Bertoldi, B.~A.~Burrington and A.~Peet,
  ``Black Holes in asymptotically Lifshitz spacetimes with arbitrary critical exponent,''
  Phys.\ Rev.\ D {\bf 80}, 126003 (2009)
  [arXiv:0905.3183 [hep-th]].

\bibitem{AyonBeato:2009nh} 
E.~Ayon-Beato, A.~Garbarz, G.~Giribet and M.~Hassaine,
``Lifshitz Black Hole in Three Dimensions,''
Phys.\ Rev.\ D {\bf 80}, 104029 (2009)
[arXiv:0909.1347 [hep-th]].


\bibitem{Dehghani:2010kd} 
  M.~H.~Dehghani and R.~B.~Mann,
  ``Lovelock-Lifshitz Black Holes,''
  JHEP {\bf 1007}, 019 (2010)
  [arXiv:1004.4397 [hep-th]].

\bibitem{Dehghani:2010gn} 
  M.~H.~Dehghani and R.~B.~Mann,
  ``Thermodynamics of Lovelock-Lifshitz Black Branes,''
  Phys.\ Rev.\ D {\bf 82}, 064019 (2010)
  [arXiv:1006.3510 [hep-th]].

\bibitem{Brenna:2011gp} 
  W.~G.~Brenna, M.~H.~Dehghani and R.~B.~Mann,
  ``Quasi-Topological Lifshitz Black Holes,''
  Phys.\ Rev.\ D {\bf 84}, 024012 (2011)
  [arXiv:1101.3476 [hep-th]].


\bibitem{Anastasiou:2013mwa}
  G.~G.~Anastasiou, M.~R.~Setare and E.~C.~Vagenas,
  ``Searching for AdS$_3$ waves and Asymptotically Lifshitz black holes in $R^3$ new massive gravity,''
  Phys.\ Rev.\ D {\bf 88} (2013) no.6,  064054
  [arXiv:1309.4704 [hep-th]].

\bibitem{Giacomini:2012hg} 
  A.~Giacomini, G.~Giribet, M.~Leston, J.~Oliva and S.~Ray,
  ``Scalar field perturbations in asymptotically Lifshitz black holes,''
  Phys.\ Rev.\ D {\bf 85}, 124001 (2012)
  [arXiv:1203.0582 [hep-th]].

\bibitem{Brenna:2015pqa} 
  W.~G.~Brenna, R.~B.~Mann and M.~Park,
  ``Mass and Thermodynamic Volume in Lifshitz Spacetimes,''
  Phys.\ Rev.\ D {\bf 92}, no. 4, 044015 (2015)
  [arXiv:1505.06331 [hep-th]].

\bibitem{Cai:2009ac} 
R.~G.~Cai, Y.~Liu and Y.~W.~Sun,
``A Lifshitz Black Hole in Four Dimensional R**2 Gravity,''
JHEP {\bf 0910}, 080 (2009)
[arXiv:0909.2807 [hep-th]].

\bibitem{AyonBeato:2010tm} 
E.~Ayon-Beato, A.~Garbarz, G.~Giribet and M.~Hassaine,
``Analytic Lifshitz black holes in higher dimensions,''
JHEP {\bf 1004}, 030 (2010)
[arXiv:1001.2361 [hep-th]].

\bibitem{Ayon-Beato:2015jga} 
E.~Ayón-Beato, M.~Bravo-Gaete, F.~Correa, M.~Hassaïne, M.~M.~Juárez-Aubry and J.~Oliva,
``First law and anisotropic Cardy formula for three-dimensional Lifshitz black holes,''
Phys.\ Rev.\ D {\bf 91}, no. 6, 064006 (2015)
[arXiv:1501.01244 [gr-qc]].

\bibitem{Balasubramanian:1999re} 
V.~Balasubramanian and P.~Kraus,
``A Stress tensor for Anti-de Sitter gravity,''
Commun.\ Math.\ Phys.\  {\bf 208}, 413 (1999)
[hep-th/9902121].

\bibitem{Brown:1986nw} 
J.~D.~Brown and M.~Henneaux,
``Central Charges in the Canonical Realization of Asymptotic Symmetries: An Example from Three-Dimensional Gravity,''
Commun.\ Math.\ Phys.\  {\bf 104}, 207 (1986).



\bibitem{Banados:1992wn} 
M.~Banados, C.~Teitelboim and J.~Zanelli,
``The Black hole in three-dimensional space-time,''
Phys.\ Rev.\ Lett.\  {\bf 69}, 1849 (1992)
[hep-th/9204099].


\bibitem{Cvetic:2001bk} 
M.~Cvetic, S.~Nojiri and S.~D.~Odintsov,
``Black hole thermodynamics and negative entropy in de Sitter and anti-de Sitter Einstein-Gauss-Bonnet gravity,''
Nucl.\ Phys.\ B {\bf 628}, 295 (2002)
[hep-th/0112045].

\bibitem{deHaro:2000vlm} 
S.~de Haro, S.~N.~Solodukhin and K.~Skenderis,
``Holographic reconstruction of space-time and renormalization in the AdS / CFT correspondence,''
Commun.\ Math.\ Phys.\  {\bf 217}, 595 (2001)
[hep-th/0002230].

\bibitem{Nutma:2013zea} 
T.~Nutma,
``xTras : A field-theory inspired xAct  package for mathematica,''
Comput.\ Phys.\ Commun.\  {\bf 185}, 1719 (2014)
[arXiv:1308.3493 [cs.SC]].

\bibitem{wip}
Ahmad Ghodsi, Farzaneh Najafi, “Work in progress”.


  \end{thebibliography}
\end{document}